\documentclass[12pt, a4paper]{article}
\usepackage[margin=2.5cm]{geometry}
\usepackage[english]{babel}
\usepackage[utf8]{inputenc}
\usepackage[T1]{fontenc}
\usepackage{titling} 

\usepackage{graphicx}
\graphicspath{{./Images/}}
\usepackage[bf]{caption} 

\usepackage{setspace}

\usepackage[square, numbers, sort&compress]{natbib} 

\usepackage{amsthm,amsmath,amssymb,bbm}
\usepackage{tensor,mathtools,slashed,braket}
\usepackage{float,wrapfig} 
\usepackage[compat=1.1.0]{tikz-feynman} 
\tikzfeynmanset{warn luatex = false} 

\usepackage{hyperref}

\newcommand{\MLC}[1]{} 
\newcommand{\email}[1]{\thanks{Electronic Address: \href{mailto:#1}{#1}}}

\begin{document}
	
	\title{Theories of the gravity+gauge type in de Sitter space}
	
	\preauthor{\begin{center}
			\large \lineskip 0.5em%
			\begin{tabular}[t]{c}}
			
			\author{Julian Lang\email{julian.lang.research@gmail.com} ~and Yasha Neiman\email{yashula@icloud.com}}
			
			\postauthor{\vspace{0.5em}\\
				\textit{Okinawa Institute of Science and Technology},\\
				\textit{1919-1 Tancha, Onna-son, Okinawa 904-0495, Japan}%
			\end{tabular}\par\end{center}}
	
	\date{\today}
	
	\begin{titlingpage}
		\maketitle
		
		\begin{abstract}
			We study theories of the ``General Relativity + Yang-Mills'' type in 4d spacetime with cosmological constant, focusing on formulations where the basic variables are connections and curvatures (but no metric). We present a new Lagrangian for GR coupled to Yang-Mills, which uses the same kind of variables for both sectors, and is suggestive of the double-copy structure. This Lagrangian comes in both non-chiral (real) and chiral (complex) versions. The chiral version makes it easy to isolate the self-dual sector. The latter lends itself to a higher-spin (HS) generalization, which combines the HS Self-Dual GR and HS Self-Dual Yang-Mills of Krasnov, Skvortsov and Tran. We then descend from the covariant formulation to lightcone gauge, and construct a cubic-exact lightcone action. For HS Self-Dual Yang-Mills, we solve the lightcone field equations to all orders in perturbation theory. Finally, we discuss the relation between lightcone foliations that share a lightray -- a setup relevant for the scattering problem in the de Sitter static patch. We derive this relation at the linearized level, and argue from causality+symmetry that it cannot receive non-linear corrections. We explore the associated gauge transformations in the non-linear covariant theory. 			
		\end{abstract}
	\end{titlingpage}
	
	\tableofcontents
	\newpage
	
	
	\onehalfspacing
	
	\section{Introduction}
	
    \subsection{Scope and motivation}
    
	This paper presents a series of results on massless theories in 4d spacetime with cosmological constant $\Lambda \neq 0$, that combine \emph{2-derivative (gravity-like)} and \emph{1-derivative (gauge-like)} interactions. The example of real-world relevance is of course General Relativity (GR) coupled to Yang-Mills (YM). In addition to GR+YM itself, we will consider its self-dual sector, as well as the self-dual sector's higher-spin (HS) generalization. 
    
    The paper originated from exploratory work on using HS Gravity \cite{Vasiliev1990} in its capacity as a working model for quantum gravity in de Sitter space \cite{Anninos2011}. Specifically, we were interested in bulk calculations of \emph{scattering in the de Sitter static patch}, in the manner of \cite{David2019,Albrychiewicz2020,Albrychiewicz2021,Neiman_2024}. Higher-Spin Gravity is a theory of infinitely many massless fields of all spins, whose interaction vertices have arbitrary numbers of derivatives (in the cubic vertices, their number is determined by the fields' spins). In this context, the gravity-like and gauge-like interactions with their low number of derivatives provide a simplified playground, closer to standard ``lower-spin'' theories. In their chiral/self-dual version, these gravity-like and gauge-like sectors constitute self-contained theories, with local \& covariant Lagrangians \cite{Krasnov2021}. The basic variables in these Lagrangians are \emph{connections and curvatures} -- similar to the master fields in Vasiliev's formalism \cite{Vasiliev1990} for full HS theory, but without the auxiliary towers of derivatives.
    
    What is remarkable about these ``connection+curvature'' formulations is that GR-like and YM-like degrees of freedom are described by the same kind of fields, and enter the Lagrangian in almost the same way. Moreover, the remaining difference in their roles is remarkably reminiscent of the ``color-kinematics duality'' or ``double-copy'' structure observed in scattering amplitudes \cite{Bern:2008qj,Bern:2010ue}, closely related to the KLT relations between closed and open strings \cite{Kawai:1985xq}. For GR+YM itself, we present a new Lagrangian formulation suggestive of these structures. In the higher-spin self-dual context, a similar formulation will allow us to combine the GR-like and YM-like theories of \cite{Krasnov2021}. In the larger context of HS Gravity, this ``unification'' of 1-derivative and 2-derivative interactions may hold some lessons about the structure of the full theory, or at least of its chiral version \cite{Ponomarev:2016lrm,Skvortsov2018,Sharapov2022a,Sharapov2022}.
    
    Meanwhile, our original project -- scattering HS fields in the static patch -- turned out \emph{not} to rely on the simplifying properties of 1- and 2-derivative interactions \cite{CubicLightconeScattering}. In particular, we ended up using the lightcone formalism of \cite{Metsaev:2018xip,Skvortsov2019}, rather than the covariant formalism of \cite{Krasnov2021}. We thus present the current paper as a standalone publication, focused on gravity-like and gauge-like interactions and their covariant formulations. Still, our list of results retains an imprint of the static-patch project, gradually flowing towards higher spins, lightcone gauge, and eventually relations between \emph{different} lightcone gauges that are relevant to de Sitter space and the static-patch problem. Our results on static-patch scattering, for full HS theory to cubic order, will be published separately \cite{CubicLightconeScattering}.
	
    \subsection{Summary and structure of the paper}
    
    The paper is structured as follows. In section \ref{sec:GR_YM}, we present our new Lagranians (chiral and non-chiral) for GR coupled to Yang-Mills, with structure suggestive of the double-copy. We explain how these Lagrangians arise from combining the Plebanski and Chalmers-Siegel formulations (of GR and Yang-Mills respectively), go over the resulting equations of motion, and show how they relate to more standard formulations. We then show how the self-dual sector of GR+YM arises as a limit of the chiral formulation. This sets the stage for section \ref{sec:covTh}, where we construct the higher-spin version of self-dual GR+YM, which combines the GR-like and YM-like theories of \cite{Krasnov2021}. Crucially, we show that the gauge symmetries of both the GR-like and YM-like sectors survive the coupling of the two sectors to each other. 
    
    In section \ref{sec:lightconegauge}, we describe a lightcone gauge, which reduces the covariant theory of section \ref{sec:covTh} to a single scalar variable per physical degree of freedom (i.e. per helicity). This generalizes the previous construction \cite{Neiman2024} to include HS self-dual YM alongside HS self-dual GR. Our lightcone gauge is more general than the lightcone formalism of \cite{Metsaev:2018xip,Skvortsov2019}, in that the leaves of our lightlike foliation can be the lightcones of bulk points, rather than boundary points. This is essential for using the lightcone formalism in de Sitter space, rather than just in Anti-de Sitter (AdS). Along the way, we point out that the lightcone field equation of HS self-dual YM, like that of self-dual YM itself, has a closed-form solution to all orders of perturbation theory, despite the presence of a cosmological constant. 
    
    In section \ref{sec:comparing_gauges}, we turn to the question of mapping between different lightcone gauges, specifically for lightlike foliations that share a lightray (but not an entire null hypersurface). As explored in \cite{Albrychiewicz2021,Neiman_2024} for YM and self-dual GR, this question is key to the kinematics of the de Sitter static-patch scattering problem (where the two lightcone gauges are adapted to the past and future horizons of the static patch). After defining the question carefully in section  \ref{sec:comparing_gauges:geometry}, we answer it using two approaches. In the first approach (section \ref{sec:comparing_gauges:field_strengths}), we determine the mapping at linear order using gauge-invariant field strengths, and then argue from causality and helicity conservation that it remains unchanged by the interactions. In \cite{CubicLightconeScattering}, we will expand this approach to full HS theory at cubic order, where causality must be demonstrated rather than assumed. In the second approach (section \ref{sec:comparing_gauges:gaugetransf}), we construct the mapping as an explicit gauge transformation within the non-linear covariant theory of section \ref{sec:covTh}. This approach is incomplete due to some unjustified assumptions, but it reproduces the correct result. Section \ref{sec:discuss} is devoted to discussion and outlook.
    
    \section{Double-copy-like Lagrangians for GR+YM} \label{sec:GR_YM}
    
    \subsection{Overview}
    
    A longtime project of theoretical physics is to make more explicit the similarities between Yang-Mills and GR. The chief modern incarnation of this project is the large literature on color-kinematics duality, or double-copy relations. While the main context of this literature is scattering amplitudes in flat spacetime \cite{Bern:2008qj,Bern:2010ue}, some work in AdS exists as well \cite{Armstrong:2020woi,Albayrak:2020fyp}. Roughly speaking, one finds that GR observables are often given by YM ones, upon a certain ``squaring'' procedure that analogizes YM gauge generators to the 4-momentum (the generator of diffeomorphisms). In recent years, a literature has also emerged that makes color-kinematics and double-copy structures manifest at the off-shell level of fields and Lagrangians. For results on free fields, see \cite{Anastasiou:2014qba,Anastasiou:2018rdx}. For general perturbative arguments at the interacting level, see \cite{Borsten:2021gyl,Borsten:2021hua,Borsten:2020zgj}. These have made more concrete for Chern-Simons theory \cite{Ben-Shahar:2021zww}, and hence for various Chern-Simons-matter theories and Chern-Simons-like formulations of YM-type theories \cite{Borsten:2023ned}, including in twistor space \cite{Borsten:2023paw,Borsten:2022vtg} and pure-spinor space \cite{Ben-Shahar:2021doh,Borsten:2023reb}. 
    
    In this section, we aim to contribute to the study of double-copy structures at the level of fields and actions, by presenting an action for GR+YM (with nonzero cosmological constant) in which both sectors are be described by the same kind of fields, and enter the Lagrangian in the same way, up to an appropriate ``squaring'' relation. Our main results (in non-chiral and chiral versions, respectively) are given in eqs. \eqref{eq:L}-\eqref{eq:V} and \eqref{eq:L_chiral}-\eqref{eq:V_chiral}.
    
    Before describing our construction, it's worth discussing the well-known ``hints'' of the double-copy in the standard Lagrangian formulation. Indeed, the simplest cartoon of the double-copy is to consider the index structure of the metric $g_{\mu\nu}$ vs. the YM potential $A_{\mu a}$. Here, it's useful to think of spacetime indices $\mu,\nu,\dots$ as spanning the generators of translations (or diffeomorphisms), while the YM indices $a,b,\dots$ span the generators of gauge rotations. We immediately see that the metric $g_{\mu\nu}$ can be ``obtained'' from $A_{\mu a}$ by trading the gauge-generator index $a$ for a second translation-generator index $\nu$. Of course, what breaks this cartoon, and makes color/kinematics duality non-trivial, is that the roles of $g_{\mu\nu}$ and $A_{\mu a}$ in the Lagrangian are quite different: for instance, $A_{\mu a}$ is a connection, while $g_{\mu\nu}$ is a tensor.
    
    At the same time, the Einstein-Yang-Mills Lagrangian contains a different hint of double-copy structure. Let us denote antisymmetric pairs of spacetime indices $[\mu_1\mu_2],[\nu_1\nu_2],\dots$ by 6d indices $I,J,\dots$, which we can think of as spanning the generators of \emph{Lorentz rotations}. Then the Riemann curvature and YM field strength can be written as $R_{IJ}$ and $F_{Ia}$, again exhibiting a color/kinematics pattern. Moreover, the Lagrangian now takes the form:
    \begin{align}
    	\mathcal{L} = \frac{1}{\kappa}R^I{}_I - \frac{1}{2g^2}F^{Ia}F_{Ia} \ , \label{eq:L_cartoon}
    \end{align}
    where $\kappa = 8\pi G$ is the GR coupling, and $g$ is the YM coupling. The two terms in \eqref{eq:L_cartoon} clearly exhibit a double-copy-like structure, with the YM term involving an extra (unavoidable) ``squaring'' step. But, of course, this is still a cartoon: we ignored the special role played by the metric in converting between curved indices $\mu,\nu,\dots$ and flat Lorentz indices $I,J,\dots$, and in providing the volume form that must multiply \eqref{eq:L_cartoon}.
    
    Our goal in this section is to present a full-fledged formulation of GR+YM, with double-copy-like structure \emph{both} in the choice of fundamental fields \emph{and} in the structure of the Lagrangian (which will be built from expressions of the general form \eqref{eq:L_cartoon}). As described in the Introduction, this work arose from the context of higher-spin theory. There, fields such as $g_{\mu\nu}$ and $A_{\mu a}$ are known as \emph{Fronsdal potentials} \cite{Fronsdal1978,Fronsdal:1978vb}. They work beautifully up to cubic vertices \cite{Sleight:2016dba}, but don't appear promising for defining the full theory. For that job, one uses ``unfolded'' formulations \cite{Vasiliev1990,Sharapov2022a,Sharapov2022}, which consist of a 1-form gauge connection and a 0-form field strength. Here, ``unfolding'' refers to the fact that these 1-forms and 0-forms include not only the physical fields, but also towers of their derivatives. In the self-dual GR-like and YM-like sectors of higher-spin theory \cite{Krasnov2021}, which are the main focus of this paper, one can write standard Lagrangians without unfolding, but still with the basic ``1-form + 0-form'' structure. In this section, we apply a similar approach to GR+YM itself.
    
    After the fact, our formulation follows from a simple (but overlooked) combination of ideas by Capovilla, Jacobson and Dell \cite{Capovilla:1989ac,Capovilla:1990qi,Capovilla:1991qb,Capovilla:1991kx}, see also Krasnov \cite{Krasnov:2011pp,Krasnov:2011up,Krasnov2016,Krasnov:2017dww}. The idea is to combine the Plebanski Lagrangian for GR and the ``Chalmers-Siegel'' Lagrangian for YM \cite{Jacobson:1987yw,Chalmers:1997sg,Chalmers:1996rq} as in \cite{Capovilla:1991qb}, and then integrate out the Plebanski 2-form as in \cite{Capovilla:1991kx}. The simplest version of the construction is chiral, which is useful for treating the full theory as a perturbation over its self-dual sector. However, in Lorentzian signature, this requires us to use complex fields, making both parity and unitarity non-manifest. Fortunately, a real, non-chiral formulation is also possible. Both versions will make use of the special properties of the Hodge dual in 4d, so the construction is specific to 4 spacetime dimensions. We will present the non-chiral version first, followed by the chiral one, followed by an account of their origin from Plebanski + Chalmers-Siegel.
    
    \subsection{Real, non-chiral Lagrangian} \label{sec:GR_YM:non_chiral}
    
    \subsubsection{Index notations}
    
    We work in a frame-like formalism, where all coordinate indices $\mu,\nu,\dots$ are subsumed into differential forms; in particular, our Lagrangians will be 4-forms. As in the vielbein formulation, we imagine a local tangent space at each point of the manifold with indices $A,B,\dots$, subject to the flat Minkowski metric $\eta_{AB}$. We will always use these indices in \emph{antisymmetric pairs}, for which we introduce composite \emph{bivector} indices $I,J,K,\ldots\equiv[A_1A_2],[B_1B_2],[C_1C_2],\ldots$, spanning the 6d space of antisymmetric $4\times 4$ matrices. This space is equipped with three structures: 
    \begin{enumerate}
    	\item A metric $\eta_{IJ}$ with $(3,3)$ signature (corresponding to spacelike and timelike bivectors). We'll use this to raise and lower indices.
    	\item A Hodge-dual, parity-odd metric $\tilde \eta_{IJ}$, also with $(3,3)$ signature.
    	\item The structure constants $f^I{}_{JK}$ of the Lorentz group.
    \end{enumerate}
    In vector indices, these structures are simply $\frac{1}{2}\eta_{A_1[B_1}\eta_{B_2]A_2}$, $\frac{1}{4}\epsilon_{A_1A_2B_1B_2}$, and $4\delta^{[A_1}_{[B_1}\eta_{B_2][C_1}\delta^{A_2]}_{C_2]}$. We denote the Hodge dual of any tensor (with respect to its 1st bivector index) as $\tilde T^{I\dots} \equiv \tilde\eta^I{}_J T^{J\dots}$. This satisfies $T_I\tilde U^I = \tilde T_I U^I$ and $\tilde{\tilde T}^I = -T^I$. 
    
    Finally, we use indices $a,b,\dots$ for the YM gauge algebra, with structure constants $f^a{}_{bc}$. These indices are trivially raised/lowered using the Killing metric $\delta_{ab}$.
    
    \subsubsection{Fields and Lagrangian}
    
    Our fundamental fields are divided into 1-forms and 0-forms. The 1-form fields are the YM connection $A^a$ and the spin-connection $\omega^I$. From these, we construct curvature 2-forms as usual:
    \begin{align}
    	F^a &\equiv dA^a + \frac{1}{2}f^a{}_{bc}A^b\wedge A^c \ ; \label{eq:F} \\ 
    	R^I &\equiv d\omega^I + \frac{1}{2}f^I{}_{JK}\omega^J\wedge\omega^K \ . \label{eq:R}
    \end{align}
    In addition, we introduce the 0-form fields $\phi^{Ia}$ and $\Psi^{IJ}$, which will be identified (through the equations of motion) with the YM curvature $F^a$ and the Weyl-curvature piece of $R^I$, respectively. The index symmetries we assume for $\Psi^{IJ}$ are:
    \begin{align}
    	\Psi^I{}_I = 0 \ ; \quad \tilde\Psi^{IJ} = \tilde\Psi^{JI} \ . \label{eq:Psi_symmetries}
    \end{align}
    These are some, but not all, of the symmetries of a Weyl tensor (in fact, they imply that the \emph{Hodge dual} $\tilde\Psi^{IJ}$ has precisely the symmetries of a Riemann tensor). The extra, non-Weyl-like components of $\Psi^{IJ}$ (i.e. the Ricci-like components of $\tilde\Psi^{IJ}$) will vanish under the equations of motion, but will contribute usefully to the variational principle. 
    
    Our Lagrangian takes the form:
    \begin{align}
    	\mathcal{L}[\omega^I,\Psi^{IJ};A^a,\phi^{Ia}] = -\frac{1}{2}M^{-1}_{IJ}\,V^I\wedge V^J \ , \label{eq:L}
    \end{align}
    where $V^I$ is a 6d ``vector'' of 2-forms, and $M_{IJ}^{-1}$ is the inverse of a 0-form matrix $M^{IJ}$. Specifically, $M^{IJ}$ and $V^I$ read:
    \begin{align}
    	M^{IJ}[\Psi^{IJ};\phi^{Ia}] &= \frac{1}{\kappa}\left(\frac{\Lambda}{3}\,\tilde\eta^{IJ} + \tilde\Psi^{IJ}\right) - \frac{1}{g^2}\,\tilde\phi_a^{(I}\phi_a^{J)} \ ; \label{eq:M} \\ 
    	V^I[\omega^I;A^a,\phi^{Ia}] &= \frac{1}{\kappa}R^I - \frac{1}{g^2}\,\phi_a^I F^a \ , \label{eq:V}
    \end{align}
    where $\kappa = 8\pi G$ is the gravitational coupling, $g$ is the YM coupling, and $\Lambda$ is the cosmological constant. Note that both of the structures \eqref{eq:M}-\eqref{eq:V} follow a double-copy-like structure similar to that of \eqref{eq:L_cartoon}. As we will see, the metric in this formalism will be contained in the 2-form $M^{-1}_{IJ}V^J$. 
    
    \subsubsection{Equations of motion} \label{sec:GR_YM:non_chiral:equations}
    
    Let us now analyze the equations of motion for the Lagrangian \eqref{eq:L}-\eqref{eq:V}, to see that it correctly describes GR+YM. The equations organize neatly in terms of the 2-form $M^{-1}_{IJ}V^J \equiv -\tilde\Sigma_I$ (the sign and Hodge dual here are for later convenience). Varying with respect to all the fields, we get the equations of motion:
    \begin{align}
    	\frac{\delta}{\delta\Psi^{IJ}} \ : \quad &\Sigma^I\wedge\Sigma^J \sim \tilde\eta^{IJ} \ ; \label{eq:vary_Psi} \\
    	\frac{\delta}{\delta\omega^I} \ : \quad &D_\omega\Sigma^I = 0 \ ; \label{eq:vary_omega} \\
    	\frac{\delta}{\delta\phi_a^I} \ : \quad &F^a\wedge\tilde\Sigma^I = \phi^a_J\,\tilde\Sigma^{(J}\wedge\Sigma^{I)} \ ; \label{eq:vary_psi} \\
    	\frac{\delta}{\delta A^a} \ : \quad &D_A(\phi^a_I\tilde\Sigma^I) = 0 \ , \label{eq:vary_A}
    \end{align}
    where $D_\omega$/$D_A$ is the exterior covariant derivative w.r.t. the corresponding connection. The undetermined component of \eqref{eq:vary_Psi} is due to the fact that, according to \eqref{eq:Psi_symmetries}, we do not vary the trace of $\Psi^{IJ}$.
    
    The constraint \eqref{eq:vary_Psi} on $\Sigma^I$ is the door through which metric geometry enters the present formalism. It has four branches of solutions \cite{DePietri:1998hnx,Freidel:1999rr} \emph{constructed from a vielbein} $e^A$, via the wedge product $(e\wedge e)^I \equiv e^{A_1}\wedge e^{A_2}$:
    \begin{align}
    	\Sigma^I = \pm (e\wedge e)^I \quad \text{or} \quad \tilde\Sigma^I = \pm (e\wedge e)^I \ . \label{eq:Sigma_e}
    \end{align}
    Regardless of the branch, eq. \eqref{eq:vary_omega} now becomes a torsion-free condition, which establishes $\omega^I$ as the spin-connection compatible with the vielbein $e^A$, and $R^I$ as its Riemann tensor. Now, let us tentatively take $\Sigma^I = (e\wedge e)^I$ as the ``correct'' branch of \eqref{eq:Sigma_e}, and plug it into the remaining equations \eqref{eq:vary_psi}-\eqref{eq:vary_A}. Eq. \eqref{eq:vary_psi} becomes:
    \begin{align}
    	F^a = \phi^a_I (e\wedge e)^I \ , \label{eq:F_psi}
    \end{align}
    which identifies the 0-forms $\phi^a_I$ as the components of the 2-form $F^a$. Eq. \eqref{eq:vary_A} becomes:
    \begin{align}
    	D_A\left(\tilde\phi^a_I (e\wedge e)^I\right) = 0 \ ,
    \end{align}
    which we now recognize via \eqref{eq:F_psi} as the YM field equation $D(\star F) = 0$. The only thing we're apparently missing is the Einstein equation. But this is actually contained in our solution $\Sigma^I = (e\wedge e)^I$ to eq. \eqref{eq:vary_Psi}! Recalling our definition $\tilde\Sigma_I = -M^{-1}_{IJ}V^J$, plugging in eqs. \eqref{eq:M}-\eqref{eq:V},\eqref{eq:F_psi} and solving for the Riemann curvature $R^I$, we have:
    \begin{align}
    	R^I = \left(\frac{\Lambda}{3}\,\eta^{IJ} + \Psi^{IJ} + \frac{\kappa}{2g^2}(\phi_a^I\phi_a^J + \tilde\phi_a^I\tilde\phi_a^J) \right)(e\wedge e)_J \ . \label{eq:Einstein}
    \end{align}
    Here, the 1st term establishes $4\Lambda$ as the Ricci scalar; the 2nd term establishes $\Psi^{IJ}$ as the Weyl curvature, while its non-Weyl-like extra components (the Ricci-like components of $\tilde\Psi^{IJ}$) are forced to vanish by the index symmetries of the Riemann tensor $R^I$; and, finally, the 3rd term establishes the traceless part of the Ricci tensor as $\kappa$ times the YM stress-energy tensor. Thus, we have precisely the Einstein equation, along with an identification of $\Psi^{IJ}$ as the Weyl curvature.
    
    Now, let's return to the other branches of solutions \eqref{eq:Sigma_e} to eq. \eqref{eq:vary_Psi}. These amount to minus signs and/or Hodge duals on the factor of $(e\wedge e)^I$ in eqs. \eqref{eq:F_psi}-\eqref{eq:Einstein}. If we pick one of the ``Hodge-dual'' branches $\tilde\Sigma^I = \pm (e\wedge e)^I$, then eq. \eqref{eq:Einstein} doesn't define a valid Riemann tensor. In particular, the $\Lambda$ term becomes a ``Hodge dual of a Ricci scalar'', which is forbidden by the Riemann's symmetries, and isn't canceled by any other term in \eqref{eq:Einstein}. Therefore, these branches don't yield solutions of the entire system. The remaining branch $\Sigma^I = -(e\wedge e)^I$ yields a valid Riemann tensor in terms of its index symmetries, but with the YM stress-energy contributing to \eqref{eq:Einstein} with the wrong  sign (in fact, at the Lagrangian level, it's the kinetic energy of the graviton that has the wrong sign in this branch \cite{Mitsou:2019nlt}). Unfortunately, we are stuck with this branch as a valid solution to the equations of motion. Our consolation is that it's discontinuously separated from the ``correct'' branch $\Sigma^I = (e\wedge e)^I$, because the $\Lambda$ term in \eqref{eq:Einstein} produces a nonzero constant value for the Ricci scalar, with opposite signs for the two branches. Overall, we see that a nonzero $\Lambda$ plays three useful roles in this formalism:
    \begin{enumerate}
    	\item It ensures that the matrix \eqref{eq:M} is invertible near the ``vacuum'' $\phi^{Ia} = \Psi^{IJ} = 0$.
    	\item It rules out the unphysical ``Hodge-dual'' branches $\tilde\Sigma^I = \pm (e\wedge e)^I$.
    	\item It ensures that the physical branch $\Sigma^I = +(e\wedge e)^I$ is discontinuous with the ghost-like branch $\Sigma^I = -(e\wedge e)^I$.
    \end{enumerate} 
    
    \subsection{Complex, chiral Lagrangian} \label{sec:GR_YM:chiral}
    
    Let us now present the chiral formulation of the same GR+YM theory. We separate the 6d space of bivectors (with indices $I,J,\dots$) into a pair of chiral, complex 3d subspaces (with indices $i,j,\dots$ and $i',j',\dots$), which we'll refer to as left-handed and right-handed, respectively. These are the eigenspaces of the Hodge dual, with eigenvalues $\mp i$. For any bivector $T^I$, the metrics $\eta_{IJ}$ and $\tilde\eta_{IJ}$ decompose as:
    \begin{align}
       \eta_{IJ}T^I T^J = \delta_{ij}T^i T^j + \delta_{i'j'}T^{i'}T^{j'} \ ; \quad \tilde\eta_{IJ}T^I T^J = i(-\delta_{ij}T^i T^j + \delta_{i'j'}T^{i'}T^{j'}) \ ,
    \end{align}
    where $\delta_{ij}$/$\delta_{i'j'}$ is the standard Euclidean metric. The Lorentz group in this language takes the form $SO(3,\mathbb{C})$, with structure constants $\sqrt{2}\,\epsilon_{ijk}$ (and their complex conjugates $\sqrt{2}\,\epsilon_{i'j'k'}$).
    
    Now, to construct a chiral formulation, we reduce our fundamental fields $(\omega^I,\Psi^{IJ},A^a,\phi^{Ia})$ to only the components along \emph{one} of the chiral subspaces. The other components are \emph{not} assumed to vanish -- they're simply no longer considered as fundamental variables. Thus, the YM connection $A^a$ and its curvature \eqref{eq:F} stay the same; instead of $\omega^I$ and $\phi^{Ia}$, we consider only $\omega^i$ and $\phi^{ia}$ (which can be seen as a rearrangement of 6 real components into 3 complex ones); instead of the full GR curvature \eqref{eq:R}, we consider:
    \begin{align}
    	R^i \equiv d\omega^i + \frac{1}{\sqrt{2}}\,\epsilon^i{}_{jk}\,\omega^j\wedge\omega^k \ ; \label{eq:R_chiral}
    \end{align}
    finally, in place of $\Psi^{IJ}$, we take the symmetric traceless matrix $\Psi^{ij}$:
    \begin{align}
    	\Psi^{ij} = \Psi^{ji} \ ; \quad \Psi^i_i = 0 \ . \label{eq:Psi_symmetries_chiral}
    \end{align}
    Note that, unlike the case of $\omega^i$ and $\phi^{ia}$, this $\Psi^{ij}$ \emph{cannot} be viewed as a mere complex rearrangement of $\Psi^{IJ}$: it genuinely has fewer components, since we removed the mixed components $\Psi^{ii'}$ from consideration. Specifically, $\Psi^{ij}$ has 5 complex components, which (together with their complex conjugates $\Psi^{i'j'}$) form precisely the 10 real components of the Weyl curvature.
    
    The chiral version of the Lagrangian \eqref{eq:L} reads:
    \begin{align}
    	\mathcal{L}[\omega^i,\Psi^{ij};A^a,\phi^{ia}] = -M^{-1}_{ij}\,V^i\wedge V^j \ , \label{eq:L_chiral}
    \end{align}
    where:
    \begin{align}
    	M^{ij}[\Psi^{ij};\phi^{ia}] &= -\frac{i}{\kappa}\left(\frac{\Lambda}{3}\,\delta^{ij} + \Psi^{ij}\right) + \frac{i}{g^2}\,\phi_a^i \phi_a^j \ ; \label{eq:M_chiral} \\ 
    	V^i[\omega^i;A^a,\phi^{ai}] &= \frac{1}{\kappa}R^i - \frac{1}{g^2}\,\phi_a^i F^a \ . \label{eq:V_chiral}   
    \end{align}
    Denoting $M^{-1}_{ij}V^j \equiv i\Sigma_i$ (so that $\Sigma_i$ will be the chiral components of $\Sigma_I$ from the previous section), the equations of motion read:
    \begin{align}
    	\frac{\delta}{\delta\Psi^{ij}} \ : \quad &\Sigma^i\wedge\Sigma^j \sim \delta^{ij} \ ; \label{eq:vary_Psi_chiral} \\
    	\frac{\delta}{\delta\omega^i} \ : \quad &D_\omega\Sigma^i = 0 \ ; \label{eq:vary_omega_chiral} \\
    	\frac{\delta}{\delta\phi_a^i} \ : \quad &F^a\wedge\Sigma^i = \phi^a_j\,\Sigma^j \wedge\Sigma^i \ ; \label{eq:vary_psi_chiral} \\
    	\frac{\delta}{\delta A^a} \ : \quad &D_A(\phi^a_i\Sigma^i) = 0 \ , \label{eq:vary_A_chiral}
    \end{align}
    The simplicity constraint \eqref{eq:vary_Psi_chiral} now has only one branch of solutions $\Sigma^i \sim (e\wedge e)^i$, in terms of a \emph{complex} vielbein (or, equivalently, a complex Urbantke metric \cite{Urbantke:1984eb}). The four real branches \eqref{eq:Sigma_e} can be expressed as $\Sigma^i = c(e\wedge e)^i$, where the vielbein is now real, and the numerical factor $c$ is $\pm 1$ or $\pm i$. The ``correct'' real branch corresponds to $c=1$. The complete basis of 2-forms $\Sigma^I$ is now spanned by $\Sigma^i$ together with their complex conjugates $\Sigma^{i'}$.
    
    With this understanding, eq. \eqref{eq:vary_omega_chiral} is again the torsion-free condition that fixes $\omega^i$ to be compatible with the vielbein. Eq. \eqref{eq:vary_psi_chiral} identifies the 0-forms $\phi^i_a$ as the chiral components of the 2-form $F^a$, as in:
    \begin{align}
    	F^a = \phi^a_i\Sigma^i + \phi^a_{i'}\Sigma^{i'} \ . \label{eq:F_psi_chiral}
    \end{align}
    Eq. \eqref{eq:vary_A_chiral} then implies the YM field equation $D(\star F) = 0$, in linear combination with the Bianchi idendity $DF = 0$. Finally, unpacking the relation $M^{-1}_{ij}V^j = i\Sigma_i$ and solving for $R^i$, we obtain the chiral analogue of \eqref{eq:Einstein}:
    \begin{align}
    	R^i = \left(\frac{\Lambda}{3}\,\delta^{ij} + \Psi^{ij} \right)\Sigma_j + \frac{\kappa}{g^2}\,\phi_a^i\phi_a^{i'}\,\Sigma_{i'} \ , \label{eq:Einstein_chiral}
    \end{align}
    which is the correct Einstein equation analogously to \eqref{eq:Einstein}, with $\Psi^{ij}$ identified as the left-handed half of the Weyl curvature. Note that \eqref{eq:Einstein_chiral} still encodes the entire Ricci part of the curvature, with its trace part $\sim \Lambda$ and traceless part $\sim \phi_a^i\phi_a^{i'}$ governed respectively by the cosmological constant and the YM stress-energy tensor. The only curvature piece missing from \eqref{eq:Einstein_chiral} is the right-handed Weyl $\Psi^{i'j'}$ (again, this doesn't mean that it vanishes, merely that it isn't captured by \eqref{eq:Einstein_chiral}; to capture it, we'd need to consider $R^{i'}$).
    
    \subsubsection{Self-dual sector} \label{sec:GR_YM:chiral:self_dual}
    
    One virtue of chiral Lagrangians such as \eqref{eq:L_chiral}-\eqref{eq:V_chiral} is that they contain the theory's self-dual sector as a limit \cite{Krasnov2016}. To arrive at the self-dual sector, we must complexify the fields, so that $\omega^{i'},\Psi^{i'j'},\phi^{i'a}$ are no longer the complex conjugates of $\omega^i,\Psi^{ij},\phi^{ia}$, but rather independent variables. We then take the limit in which the left-handed field strengths $\Psi^{ij}$ and $\phi^{ia}$ are small. In other words, we expand the Lagrangian \eqref{eq:L_chiral}-\eqref{eq:V_chiral} in powers of $\Psi^{ij}$ and $\phi^{ia}$. The zeroth-order term $-\frac{3i}{\kappa\Lambda}R_i\wedge R^i$ is topological, and can be discarded. The self-dual sector is then contained in the first-order terms:
    \begin{align}
       \mathcal{L}_{\text{self-dual}}[\omega^i,\Psi^{ij};A^a,\phi^{ia}] = \frac{9i}{\kappa\Lambda^2}\,\Psi_{ij}R^i\wedge R^j + \frac{6i}{g^2\Lambda}\,\phi_{ia}R^i\wedge F^a \ . \label{eq:L_SD}
    \end{align}
    This Lagrangian describes right-handed gravitational and YM fields, encoded in $\omega^i$ and $A^a$, along with linearized left-handed perturbations, encoded in $\Psi^{ij}$ and $\phi^{ia}$ (note that, somewhat confusingly, the \emph{left-handed} spin-connection $\omega^i$ is encoding \emph{right-handed} degrees of freedom). In this limit, the left-handed perturbations don't backreact on the nonlinear right-handed fields. In particular, this implies that the YM fields don't backreact on the geometry, since the YM stress-energy tensor contains a factor of $\phi^{ia}$ -- see \eqref{eq:Einstein_chiral}. Moreover, we see from \eqref{eq:Einstein_chiral} that the curvature 2-form $R^i$ now coincides with the metric-like 2-form $\Sigma^i$, up to a constant factor of $\Lambda/3$.
    
    \subsection{Derivation from a Plebanski-type Lagrangian}
    
    Having analyzed the new Lagrangians \eqref{eq:L},\eqref{eq:L_chiral} on their own terms, let us now demystify them by ``integrating in'' the 2-forms $\Sigma$ as independent variables. In the non-chiral case, this gives the Lagrangian:
    \begin{align}
       \mathcal{L}[\Sigma^I,\omega^I;\Psi^{IJ},A^a,\phi_a^I] &= \frac{1}{2}M_{IJ}\tilde\Sigma^I\!\wedge\tilde\Sigma^J + V_I\wedge\tilde\Sigma^I \label{eq:L_unpacked} \\
       	  &= \frac{1}{\kappa}\,\tilde\Sigma^I \wedge\left(R_I - \frac{\Lambda}{6}\,\Sigma_I - \frac{1}{2}\Psi_{JI}\Sigma^J \right) - \frac{1}{g^2}\,\phi^a_I\,\tilde\Sigma^I \wedge\left(F^a - \frac{1}{2}\phi_J^a\,\Sigma^J \right) \ . \nonumber
    \end{align}
    Varying this w.r.t. $\Sigma^I$, we obtain $\tilde\Sigma_I = -M_{IJ}^{-1}V^J$ as a field equation. Plugging this back into the Lagrangian to eliminate $\Sigma^I$, we recover \eqref{eq:L}. The chiral Lagrangian \eqref{eq:L_chiral} arises in the same way from:
    \begin{align}
        \mathcal{L}[\Sigma^i,\omega^i;\Psi^{ij},A^a,\phi_a^i] &= -M_{ij}\Sigma^i\wedge\Sigma^j + 2iV_i\wedge\Sigma^j \label{eq:L_unpacked_chiral} \\
    	  &= \frac{2i}{\kappa}\,\Sigma^i\wedge \left(R_i - \frac{\Lambda}{6}\,\Sigma_i - \frac{1}{2}\Psi_{ij}\Sigma^j \right) - \frac{2i}{g^2}\,\phi_i^a\Sigma^i\wedge \left( F_a - \frac{1}{2}\phi^a_j\Sigma^j \right) \ . \nonumber
    \end{align}
    Unlike our main Lagrangians \eqref{eq:L},\eqref{eq:L_chiral}, the unpacked versions \eqref{eq:L_unpacked},\eqref{eq:L_unpacked_chiral} take the standard form ``GR term + YM term'' (as does the Lagrangian \eqref{eq:L_SD} of the self-dual theory). In particular, the GR term in \eqref{eq:L_unpacked_chiral} is just the Plebanski Lagrangian, while the YM term is the Chalmers-Siegel Lagrangian (adapted to curved spacetime, as in \cite{Capovilla:1991qb}). The corresponding terms in \eqref{eq:L_unpacked} are just the non-chiral versions of these.
     
	\section{Higher-spin self-dual GR+YM} \label{sec:covTh}
	
	\subsection{Notations, fields and Lagrangian}
	
	We now turn to the paper's main focus: the higher-spin extension of the self-dual GR+YM theory \eqref{eq:L_SD}, which will combine the higher-spin self-dual GR and YM theories of \cite{Krasnov2021}. We continue to use differential-form notation on the 4d spacetime, just like in \eqref{eq:L_SD}. The YM color structure, described by the indices $(a,b,\dots)$, also remains the same. The higher-spin fields will be characterized by increasing numbers of the internal left-handed bivector indices $(i,j,\dots)$. Thus, the 1-form $\omega^i$ and 0-form $\Psi^{ij}$ from the spin-2 GR sector generalize to spin $s$ as $\omega^{i_1\dots i_{s-1}}$ and $\Psi^{i_1\dots i_s}$. Similarly, the spin-1 YM fields $A_a,\phi_a^i$ generalize as $A_a^{i_1\dots i_{s-1}}$ and $\phi_a^{i_1\dots i_s}$. The fields will all be totally symmetric and traceless w.r.t. the bivector indices $(i,j,\dots)$, as we already saw with the left-handed spin-2 curvature $\Psi^{ij}$. It is then very convenient to convert every bivector index into a pair of left-handed (2-component) spinor indices $(\alpha,\beta,\dots)$. Total symmetry and tracelessness w.r.t. $(i,j,\dots)$ is then simply total symmetry w.r.t. the spinor indices. In particular, the spin-2 GR fields become $\omega^{\alpha\beta}$ and $\Psi^{\alpha\beta\gamma\delta}$, while the spin-1 YM fields become $A_a$ and $\phi_a^{\alpha\beta}$, with all spinor indices totally symmetric. 
	
	We raise and lower spinor indices using the $2\times 2$ antisymmetric matrix $\epsilon_{\alpha\beta}$:
	\begin{align}
	   \zeta_\alpha = \epsilon_{\alpha\beta}\zeta^\beta \ ; \quad \zeta^\alpha = \zeta_\beta\epsilon^{\beta\alpha} \ . \label{eq:spinor_indices}
	\end{align}
	As a final notational touch, we feed the fields of all the different spins into generating functions w.r.t. a spinor coordinate $y^\alpha$. Thus, the GR-like and YM-like fields of all spins are now packaged into functions of both $y^\alpha$ and spacetime position $x^\mu$:
	\begin{align}
	  \omega(x;y) &= \sum_s\frac{y^{\alpha_1}\ldots y^{\alpha_{2s-2}}}{(2s-2)!}\,\omega_{\alpha_1\ldots\alpha_{2s-2}}(x) \ ; \quad
	  \Psi(x;y) = \sum_s\frac{y^{\alpha_1}\ldots y^{\alpha_{2s}}}{(2s)!}\,\Psi_{\alpha_1\ldots\alpha_{2s}}(x) \ ; \label{eq:GR_like_fields} \\
	  A^a(x;y) &= \sum_s\frac{y^{\alpha_1}\ldots y^{\alpha_{2s-2}}}{(2s-2)!}\,A^a_{\alpha_1\ldots\alpha_{2s-2}}(x) \ ; \quad
      \phi^a(x;y) = \sum_s\frac{y^{\alpha_1}\ldots y^{\alpha_{2s}}}{(2s)!}\,\phi^a_{\alpha_1\ldots\alpha_{2s}}(x) \ , \label{eq:YM_like_fields}
	\end{align}
	where $\omega \equiv \omega_\mu dx^\mu$ and $A^a \equiv A^a_\mu dx^\mu$ are again 1-forms w.r.t. the spacetime coordinates $x^\mu$, while $\Psi,\phi_a$ are 0-forms. We can restrict the GR-like fields \eqref{eq:GR_like_fields} to \emph{even} nonzero spins $s$, and the YM-like fields \eqref{eq:YM_like_fields} to \emph{odd} spins. Beyond this, the only possible restrictions are:
	\begin{itemize}
		\item Restrict the GR-like fields to $s=2$, which gives higher-spin self-dual YM coupled to standard self-dual GR.
		\item Restrict the GR-like fields to $s=2$ \emph{and} restrict the YM-like fields to $s=1$, which takes us back to standard self-dual GR+YM, described by \eqref{eq:L_SD}.
    \end{itemize}
	As in \eqref{eq:L_SD}, the connection 1-forms $\omega,A_a$ will encode non-linearly interacting right-handed degrees of freedom, while the 0-forms $\Psi,\phi_a$ encode linearized left-handed degrees of freedom propagating on top of $\omega,A_a$. 
	
	The left-handed Lorentz commutator in \eqref{eq:R_chiral} gets replaced by its higher-spin generalization, which is simply a Poisson bracket w.r.t. the spinor $y^\alpha$:
    \begin{align}
    	\left\{f(y),g(y)\right\} \equiv \frac{1}{2}\,\partial^\alpha f\partial_\alpha g \ ; \quad \partial_\alpha \equiv \frac{\partial}{\partial y^\alpha} \ . \label{eq:Poisson_bracket}
	\end{align}
	When one or both of the functions $f,g$ are quadratic in $y^\alpha$, the bracket \eqref{eq:Poisson_bracket} reduces back to the left-handed Lorentz commutator. Using \eqref{eq:Poisson_bracket} and the usual YM gauge algebra, we can construct 2-form curvatures $R \equiv \frac{1}{2}R_{\mu\nu}dx^\mu\wedge dx^\nu$ and $F^a \equiv \frac{1}{2}F^a_{\mu\nu}dx^\mu\wedge dx^\nu$ out of the connections $\omega$ and $A_a$: 
	\begin{align}
		R &= d\omega + \frac{1}{2}\{\omega,\omega\} \ ; \label{eq:HS_R} \\
		F^a &= dA^a + \{\omega,A^a\} + \frac{1}{2}f^a{}_{bc}A^b\wedge A^c \ , \label{eq:HS_F}
	\end{align}
	Eq. \eqref{eq:HS_R} reduces to the left-handed Lorentz curvature \eqref{eq:R_chiral} when $\omega$ is purely spin-2; similarly, \eqref{eq:HS_F} reduces to \eqref{eq:F} when $A_a$ is purely spin-1. Note that the second term in \eqref{eq:HS_F} is non-vanishing only when $A^a$ contains higher spins; its purpose is to make $F_a$ covariant w.r.t. the GR-like connection $\omega$.
	
   The notations above are sufficient to write the equations of motion for the right-handed degrees of freedom:
	\begin{align}
    	R\wedge R &= 0 \ ; \label{eq:R_R_EOM} \\ 
    	R\wedge F^a &= 0 \ . \label{eq:R_F_EOM}
   \end{align}
   To incorporate these into a Lagrangian, we need to introduce the left-handed degrees of freedom $\Psi,\phi^a$ as Lagrange multipliers. To write down the appropriate products, we introduce notations for contracting the spinor indices inside the Taylor coefficients of two spinor functions $f(y),g(y)$:
   \begin{align}
		f(y)\triangleright g(y) &\equiv f\!\left(\frac{\partial}{\partial y}\right) g(y) \ ; \label{eq:contraction} \\ 
		\Braket{f|g} &\equiv \big(f(y)\triangleright g(y)\big)\Big|_{y=0} \ . \label{eq:pairing}
    \end{align}
    Note that the product \eqref{eq:contraction} is the dual of ordinary multiplication $f(y)g(y)$ under the pairing \eqref{eq:pairing}.
     
	We are now ready to present the Lagrangian for higher-spin self-dual GR+YM, which straightforwardly generalizes \eqref{eq:L_SD}:
	\begin{align}
	   \mathcal{L}[\omega,\Psi;A^a,\phi^a] = \frac{i}{2}\Braket{\Psi|R\wedge R} + i\Braket{\phi_a|R\wedge F^a} \ . \label{eq:HS_L}
	\end{align}
	Here, we subsumed any numerical factors into the definition of the linearized left-handed fields $\Psi$ and $\phi_a$. The first term in \eqref{eq:HS_L} is just the HS self-dual GR Lagrangian from \cite{Krasnov2021}. The second term is the HS self-dual YM Lagrangian of \cite{Krasnov2021}, with the dynamical $R(x,y)$ in place of a fixed spin-2 geometric background. The non-trivial statement in writing the combined Lagrangian \eqref{eq:HS_L} is that the gauge symmetries discussed in \cite{Krasnov2021} survive the coupling of the GR-like and YM-like sectors to each other. We will demonstrate this in section \ref{sec:covTh:proof} below.
	
	\subsection{Gauge symmetries}
	
    The gauge symmetries of the Lagrangian \eqref{eq:HS_L} are parameterized by the following quantities \cite{Krasnov2021}:
	\begin{align}
		\Theta(x;y) &= \sum_s\frac{y^{\alpha_1}\ldots y^{\alpha_{2s-2}}}{(2s-2)!}\,\Theta_{\alpha_1\ldots\alpha_{2s-2}}(x) \ ; \label{eq:Theta} \\
		\Xi^\mu(x;y) &= \sum_s\frac{y^{\alpha_1}\ldots y^{\alpha_{2s-4}}}{(2s-4)!}\,\Xi^\mu_{\alpha_1\ldots\alpha_{2s-4}}(x) \ ; \label{eq:Xi} \\
		\theta^a(x;y) &= \sum_s\frac{y^{\alpha_1}\ldots y^{\alpha_{2s-2}}}{(2s-2)!}\,\theta^a_{\alpha_1\ldots\alpha_{2s-2}}(x) \ ; \label{eq:theta} \\
        \xi^{\mu a}(x;y) &= \sum_s\frac{y^{\alpha_1}\ldots y^{\alpha_{2s-4}}}{(2s-4)!}\,\xi^{\mu a}_{\alpha_1\ldots\alpha_{2s-4}}(x) \ . \label{eq:xi} 
	\end{align}
	The parameters \eqref{eq:Theta}-\eqref{eq:Xi} belong to the GR-like sector, and are thus uncolored, with even nonzero spins $s$. At $s=2$, $\Theta^{\alpha\beta}(x)$ parameterizes left-handed Lorentz rotations, while $\Xi^\mu(x)$ paremeterizes diffeomorphisms; as we can see in \eqref{eq:Theta}-\eqref{eq:Xi}, both of these acquire HS generalizations. The other two gauge parameters \eqref{eq:theta}-\eqref{eq:xi} belong to the YM-like sector, and are therefore colored, with odd spins $s$. At $s=1$, $\theta^a(x)$ is the standard YM gauge parameter. The ``colored diffeomorphisms'' $\xi_a^\mu$ are a purely higher-spin phenomenon, appearing only at $s>1$.  
	
	To specify the transformations of the fields \eqref{eq:GR_like_fields}-\eqref{eq:YM_like_fields} under these gauge parameters \eqref{eq:Theta}-\eqref{eq:xi}, we'll need to introduce another product on spinor functions, which is the dual of the Poisson bracket \eqref{eq:Poisson_bracket} under the pairing \eqref{eq:pairing}:
	\begin{align}
		f(y)\circ g(y) &\equiv \frac{y^\alpha}{2}\Big((\partial_\alpha f)\triangleright g\Big) \ . \label{eq:circ}
    \end{align}
    We now define the gauge transformations as:
    \begin{align}
	  \delta_\Theta\omega_\mu &= \partial_\mu\Theta + \{\omega_\mu,\Theta\} ; & \delta_\Theta\Psi &= \Theta\circ\Psi \ ; \label{eq:Theta_GR} \\
	  \delta_\Theta A_\mu^a &= \{A_\mu^a,\Theta\} \ ; & \delta_\Theta\phi^a &= \Theta\circ\phi^a \ ; \label{eq:Theta_YM} \\
      \delta_\theta\omega_\mu &= 0 \ ; & \delta_\theta\Psi &= \theta_a\circ\phi^a \ ; \label{eq:theta_GR} \\
      \delta_\theta A_\mu^a &= \partial_\mu\theta^a + \{\omega_\mu,\theta^a\} + f^a{}_{bc} A_\mu^b\theta^c \ ; & \delta_\theta \phi^a &= f^a{}_{bc} \phi^b\theta^c \ ; \label{eq:theta_YM} \\
      \delta_\Xi\omega_\mu &= \Xi^\nu R_{\nu\mu} \ ; & \delta_\Xi\Psi &= \Xi^\mu\triangleright\big(\partial_\mu\Psi - \omega_\mu\circ\Psi + \{\phi_a,A_\mu^a\} \big) \ ; \label{eq:Xi_GR} \\
      \delta_\Xi A_\mu^a &= \Xi^\nu F^a_{\nu\mu} \ ; & \delta_\Xi\phi^a &= \Xi^\mu\triangleright (\partial_\mu\phi^a - \omega_\mu\circ\phi^a + f^a{}_{bc}A_\mu^b\phi^c) \ ; \label{eq:Xi_YM} \\
      \delta_\xi\omega_\mu &= 0 \ ; & \delta_\xi\Psi &= \xi_a^\mu\triangleright (\partial_\mu\phi^a - \omega_\mu\circ\phi^a + f^a{}_{bc}A_\mu^b\phi^c) \ ; \label{eq:xi_GR} \\
      \delta_\xi A_\mu^a &= \xi^{a\nu} R_{\nu\mu} \ ; & \delta_\xi\phi^a &= 0 \ . \label{eq:xi_YM}
   \end{align}
   Here, we made the spacetime indices on differential forms explicit, to show how they contract with the diffeomorphism generators $\Xi^\mu,\xi^\mu_a$. When $\Theta$ is purely spin-2, eqs. \eqref{eq:Theta_GR}-\eqref{eq:Theta_YM} reduce to left-handed Lorentz rotations. Similarly, when $\theta^a$ is purely spin-1, eqs. \eqref{eq:theta_GR}-\eqref{eq:theta_YM} reduce to a standard YM gauge transformation. Finally, when $\Xi^\mu$ is purely spin-2, eqs. \eqref{eq:Xi_GR}-\eqref{eq:Xi_YM} reduce to a standard diffeomorphism, covariantized w.r.t. the connections $\omega,A^a$. 
   
   Compared with \cite{Krasnov2021}, the transformations \eqref{eq:Theta_GR}-\eqref{eq:xi_YM} contain some new terms, which express the coupling between the GR-like and YM-like sectors. The non-trivial additions are in the transformation rules for $\Psi$, specifically $\delta_\theta\Psi$, $\delta_\xi\Psi$ and the last term in $\delta_\Xi\Psi$. We engineered these to ensure the overall invariance of the Lagrangian, as we'll see below. 
	
   \subsection{Hierarchy of fields-on-backgrounds} \label{sec:covTh:hierarchy}
   
   Before we prove the gauge invariance of the higher-spin self-dual Lagrangian \eqref{eq:HS_L}, we pause to discuss the ways in which some of its fields act as an inert background for others. First, as already discussed, the left-handed degrees of freedom $\Psi,\phi^a$ behave as linearized perturbations over the right-handed background defined by $\omega,A^a$ (we revert to index-free notation for differential forms). In particular, the field equations \eqref{eq:R_R_EOM}-\eqref{eq:R_F_EOM} that govern $\omega,A^a$ are independent of $\Psi,\phi^a$. 
   
   Second, within the right-handed sector, the GR-like connection $\omega$ acts as a background for the YM-like connection $A^a$. This manifests in two ways:
   \begin{enumerate}
     \item The field equation \eqref{eq:R_R_EOM} that governs $\omega$ is independent of $A^a$.
     \item As we can see in \eqref{eq:theta_GR},\eqref{eq:xi_GR}, $\omega$ is not affected by the YM-like sector of gauge transformations.
   \end{enumerate}
   Finally, the low-spin components of the right-handed sector act as backgrounds for the higher-spin fields. Indeed, as noted in \cite{Neiman2024}, the following holds for the $s=2$ spin connection $\omega_{\alpha\beta}(x)$:
   \begin{enumerate}
 	\item The field equation that governs the spin-2 $\omega_{\alpha\beta}$ is the 4th-order-in-$y^\alpha$ piece of \eqref{eq:R_R_EOM}. This is the usual field equation of self-dual GR, without any involvement from higher-spin fields.
 	\item The spin-2 $\omega_{\alpha\beta}$ is unaffected by the $s>2$ components of the gauge transformations \eqref{eq:Theta_GR},\eqref{eq:Xi_GR}, while their $s=2$ component acts as in standard GR.
   \end{enumerate}
   Similarly, we notice the following for the $s=1$ YM connection $A^a(x)$:
   \begin{enumerate}
   	\item The field equation that governs the spin-1 $A^a$ is the 2nd-order-in-$y^\alpha$ piece of \eqref{eq:R_F_EOM}. This is the usual field equation of self-dual YM on the self-dual GR background defined by the spin-2 $\omega_{\alpha\beta}$, without any involvement from higher-spin fields.
   	\item The spin-1 $A^a$ is unaffected by the $s>2$ components of the GR-like gauge transformations \eqref{eq:Theta_YM},\eqref{eq:Xi_YM}, while their $s=2$ component acts as in standard GR. 
   	\item The spin-1 $A^a$ is unaffected by the ``colored diffeomorphisms'' \eqref{eq:xi_YM}, or by $s>1$ components of the YM-like gauge transformations \eqref{eq:theta_YM}, while the $s=1$ component of \eqref{eq:theta_YM} acts as in standard YM.
   \end{enumerate} 

	\subsection{Proof of gauge invariance} \label{sec:covTh:proof}
	
    \subsubsection{Rearranging the left-handed degrees of freedom}
   
    Let us now demonstrate that the transformations \eqref{eq:Theta_GR}-\eqref{eq:xi_YM} are indeed symmetries of the theory \eqref{eq:HS_L}. This will become easier after a slight repackaging of the left-handed degrees of freedom $\Psi,\phi^a$. Specifically, we Fourier-transform $\Psi$ and $\phi^a$ with respect to the spinor $y^\alpha$ \cite{Neiman2024}:
    \begin{align}
    	 \tilde\Psi(x;y) \equiv \int\frac{d^2u}{(2\pi)^2}\,e^{u_\alpha y^\alpha}\,\Psi(x;u) \ ; \quad \tilde\phi^a(x;y) \equiv \int\frac{d^2u}{(2\pi)^2}\,e^{u_\alpha y^\alpha}\,\phi^a(x;u) \ . \label{eq:Fourier}
    \end{align}
    Note that for any fixed spins in $\Psi$ or $\phi^a$, the Fourier-transformed fields \eqref{eq:Fourier} are distributional w.r.t. $y^\alpha$. With the new fields \eqref{eq:Fourier}, the convolutional products \eqref{eq:contraction}-\eqref{eq:pairing},\eqref{eq:circ} are no longer necessary; all operations become local in $y^\alpha$, just as they are local in $x^\mu$. The Lagrangian \eqref{eq:HS_L} becomes an integral over spinor space, reminiscent of superfield Lagrangians (though our spinor coordinates $y^\alpha$ are of course commuting):
    \begin{align}
   	   \mathcal{L}(x) &= i\int d^2y\,L(x;y) \ ; \label{eq:L_integral} \\ 
   	   L(x;y) &= \frac{1}{2}\tilde\Psi R\wedge R + \tilde\phi_a R\wedge F^a \ . \label{eq:L_integrand}
    \end{align}
    Here, we highlighted the Lagrangian's dependence on the coordinates $x^\mu,y^\alpha$, rather than its dependence on the dynamical fields $\omega,\tilde\Psi,A^a,\tilde\phi^a$. We pushed the overall factor of $i$ into the definition \eqref{eq:L_integral} for convenience. The gauge transformations \eqref{eq:Theta_GR}-\eqref{eq:xi_YM} can now be expressed compactly as:
    \begin{gather}
	  \delta_\Theta\omega = D\Theta \ ; \quad \delta_\Theta\tilde\Psi = \{\tilde\Psi,\Theta\} \ ; \quad \delta_\Theta A^a = \{A^a,\Theta\} \ ; \quad \delta_\Theta\tilde\phi^a = \{\tilde\phi^a,\Theta\} \ ; \label{eq:Theta_transformations} \\
      \delta_\theta\omega = 0 \ ; \quad \delta_\theta\tilde\Psi = \left\{\tilde\phi_a,\theta^a\right\} \ ; \quad \delta_\theta A^a = D\theta^a \ ; \quad \delta_\theta\tilde\phi^a = f^a{}_{bc}\tilde\phi^b\theta^c \ ; \label{eq:theta_transformations} \\
      \delta_\Xi\omega = \Xi\,\lrcorner\,R \ ; \quad \delta_\Xi\tilde\Psi = \Xi\,\lrcorner\,\big(D\tilde\Psi + \{\phi_a,A^a\}\big) \ ; \quad
      \delta_\Xi A^a = \Xi\,\lrcorner\,F^a \ ; \quad \delta_\Xi\tilde\phi^a = \Xi\,\lrcorner\,D\tilde\phi^a \ ; \label{eq:Xi_transformations} \\
      \delta_\xi\omega = 0 \ ; \quad \delta_\xi\tilde\Psi = \xi_a\,\lrcorner\,D\tilde\phi^a \ ; \quad \delta_\xi A^a = \xi^a\,\lrcorner\,R \ ; \quad \delta_\xi\tilde\phi^a = 0 \ . \label{eq:xi_transformations}
   	\end{gather}
   	Here, we again use index-free notation for differential forms, and use the ``$\lrcorner$'' symbol for their interior product by the diffeomorphism vectors $\Xi^\mu,\xi_a^\mu$. The covariant exterior derivative $D$ is defined on uncolored/colored objects $v$/$v^a$ as:
   	\begin{align}
   		Dv \equiv dv + \{\omega,v\} \ ; \quad Dv^a \equiv dv^a + \{\omega,v^a\} + f^a{}_{bc}A^b\wedge v^c \label{eq:D}
    \end{align}
    
    \subsubsection{Useful identities}
    
    We now collect some identities that will be useful for the gauge invariance analysis. First, we see from \eqref{eq:HS_R}-\eqref{eq:HS_F},\eqref{eq:D} that the covariant exterior derivatives \eqref{eq:D} square as:
    \begin{align}
    	D^2 v = \{R,v\} \ ; \quad D^2 v^a = \{R,v^a\} + f^a{}_{bc}F^b\wedge v^c \ . \label{eq:D_squared}
    \end{align}
    Second, we see that under any infinitesimal variation $\delta\omega,\delta A^a$ of the connections $\omega,A^a$, the variation of the 2-forms $R,F^a$ reads:
    \begin{align}
    	\delta R = D(\delta\omega) \ ; \quad \delta F^a = D(\delta A^a) + \{\delta\omega, A^a\} \ . \label{eq:delta_R_F}
    \end{align}
    Third, the covariant derivatives of $R$ and $F^a$ read:
    \begin{align}
	  DR = 0 \ ; \quad DF^a = \{R, A^a\} \ . \label{eq:D_R_F}
    \end{align}    
    Fourth, for any 0-form $f$, 4-form $g$ and vector $\Xi^\mu$, we have the following identity, reminiscent of the Leibnitz rule:
    \begin{align}
    	D\big(\Xi\,\lrcorner\,(fg)\big) = (\Xi\,\lrcorner\,Df)g + fD(\Xi\,\lrcorner\,g) \ . \label{eq:D_Xi}
    \end{align}
    To prove this, we note that for \emph{general} forms $f$ and $g$, both the exterior derivative $D(\dots)$ and the interior product $\Xi\,\lrcorner\,(\dots)$ act as \emph{anti-derivations}, i.e. they satisfy a graded Leibnitz rule:
    \begin{align}
    	\begin{split}
       	  D(f\wedge g) &= Df\wedge g + (-1)^p f\wedge Dg \ ; \\
       	  \Xi\,\lrcorner\,(f\wedge g) &= (\Xi\,\lrcorner\,f)\wedge g + (-1)^p f\wedge(\Xi\,\lrcorner\,g) \ ,
    	\end{split} \label{eq:anti_derivations}
    \end{align}
    where $p$ is the rank of $f$. From \eqref{eq:anti_derivations}, it follows that the \emph{anticommutator} of $D(\dots)$ and $\Xi\,\lrcorner\,(\dots)$ is itself a \emph{derivation}, i.e. it satisfies the Leibnitz rule:
    \begin{align}
	  D\big(\Xi\,\lrcorner\,(f\wedge g)\big) + \Xi\,\lrcorner\,D(f\wedge g) = \big(D(\Xi\,\lrcorner\,f) + \Xi\,\lrcorner\,Df\big)\wedge g + f\wedge\big(D(\Xi\,\lrcorner\,g) + \Xi\,\lrcorner\,Dg \big) \ . \label{eq:derivation}
    \end{align}
    We now specialize back to a 0-form $f$ and 4-form $g$. Since 0-forms have no interior products, and 4-forms have no exterior derivatives, some of the terms in \eqref{eq:derivation} vanish, and we arrive at the desired identity \eqref{eq:D_Xi}. In the exact same way, if we replace $D(\dots)$ by the Poisson bracket with a 1-form $\{\omega,\dots\}$, we get the identity:
    \begin{align}
    	\big\{\omega, \Xi\,\lrcorner\,(fg)\big\} = \big(\Xi\,\lrcorner\,\{\omega,f\}\big)g + f\{\omega,\Xi\,\lrcorner\,g\} \ . \label{eq:omega_Xi}
    \end{align}
    Finally, the identities \eqref{eq:D_Xi},\eqref{eq:omega_Xi} maintain their form if we add color indices to any of the objects.
    
	\subsubsection{Gauge variations of the Lagrangian}
	
    Let us now analyze the variations of the Lagrangian density \eqref{eq:L_integrand} under the gauge transformations \eqref{eq:Theta_transformations}-\eqref{eq:xi_transformations}. Note that the following variations of $L$ are equivalent to zero, i.e. signify a symmetry:
    \begin{enumerate}
    	\item As usual, any exterior derivative $d(\dots)$, since this will vanish when integrated over spacetime.
    	\item Any Poisson bracket $\{f,g\}$, since this can be written as a total divergence $\frac{1}{2}\partial^\alpha(f\partial_\alpha g)$ in spinor space, and will vanish when integrated in \eqref{eq:L_integral}. 
    	\item Any covariant exterior derivative $D(\dots)$, since this is a linear combination of the previous two cases.
    \end{enumerate}
    We will denote expressions equivalent to zero in this sense by $\approx 0$.
    
    Now, consider the HS left-handed Lorentz rotations \eqref{eq:Theta_transformations}. From \eqref{eq:Theta_transformations},\eqref{eq:D}-\eqref{eq:delta_R_F}, we conclude that the 2-form and 0-form field strengths $R,\tilde\Psi,F^a,\tilde\phi^a$ all transform covariantly (for $F^a$ this requires some calculation):
    \begin{align}
    	\delta_\Theta R = \{R,\Theta\} \ ; \quad \delta_\Theta\tilde\Psi = \{\tilde\Psi,\Theta\} \ ; \quad \delta_\Theta F^a = \{F^a,\Theta\} \ ; \quad \delta_\Theta\tilde\phi^a = \{\tilde\phi^a,\Theta\}
    \end{align}
    Therefore, the same is true of (both terms in) the Lagrangian integrand \eqref{eq:L_integrand}, and we get:
    \begin{align}
    	\delta_\Theta L = \{L,\Theta\} \approx 0 \ . \label{eq:Theta_L}
    \end{align}
    Thus, \eqref{eq:Theta_transformations} is indeed a symmetry of the theory. Let us move on to the HS color rotations \eqref{eq:theta_transformations}. We see from \eqref{eq:theta_transformations},\eqref{eq:D_squared}-\eqref{eq:delta_R_F} that the field strengths transform as:
    \begin{align}
    	\delta_\theta R = 0 \ ; \quad \delta_\theta\tilde\Psi = \left\{\tilde\phi_a,\theta^a\right\} \ ; \quad \delta_\theta F^a = \{R,\theta^a\} + f^a{}_{bc}F^b\theta^c \ ; \quad \delta_\theta\tilde\phi^a = f^a{}_{bc}\tilde\phi^b\theta^c \ .
    \end{align}
    If $\delta_\theta\tilde\Psi$ and the first term in $\delta_\theta F^a$ were absent, these would be the standard convariant transformations, and both terms in $L$ would be invariant. Due to the extra transformation terms, the two terms in $L$ receive non-trivial variations, which combine as:
    \begin{align}
      \begin{split}
    	\delta_\theta L &= \frac{1}{2}\delta_\theta\left(\tilde\Psi R\wedge R\right) + \delta_\theta\left(\tilde\phi_a R\wedge F^a\right) = \frac{1}{2}\left\{\tilde\phi_a,\theta^a\right\}R\wedge R + \tilde\phi_a R\wedge \{R,\theta^a\} \\
    	  &= \frac{1}{2}\left\{\tilde\phi_a R\wedge R,\theta^a\right\} \approx 0 \ ,
      \end{split}
    \end{align}
    again giving a symmetry. Now, consider the HS diffeomorphisms \eqref{eq:Xi_transformations}. From \eqref{eq:Xi_transformations},\eqref{eq:delta_R_F}, we see that the field strengths transform as:
    \begin{align}
      \delta_\Xi R &= D(\Xi\,\lrcorner\,R) \ ; & \delta_\Xi\tilde\Psi &= \Xi\,\lrcorner\,\big(D\tilde\Psi + \{\phi_a,A^a\}\big) \ ; \\ 
      \delta_\Xi F^a &= D(\Xi\,\lrcorner\,F^a) + \{\Xi\,\lrcorner\,R,A^a\} \ ; & \delta_\Xi\tilde\phi^a &= \Xi\,\lrcorner\,D\tilde\phi^a \ .
    \end{align}
    Using this and \eqref{eq:D_R_F},\eqref{eq:anti_derivations}, we find that the two terms in $L$ transform as:
    \begin{align}
      \frac{1}{2}\delta_\Xi\left(\tilde\Psi R\wedge R\right) &= \frac{1}{2}\left(\Xi\,\lrcorner\,\big(D\tilde\Psi + \{\phi_a,A^a\}\big)\right)R\wedge R + \frac{1}{2}\tilde\Psi D\big(\Xi\,\lrcorner\,(R\wedge R)\big) \ ; \\
      \delta_\Xi\left(\tilde\phi_a R\wedge F^a\right) &= \left(\Xi\,\lrcorner\,D\tilde\phi_a\right)R\wedge F^a + \tilde\phi_a \left(D\big(\Xi\,\lrcorner\,(R\wedge F^a)\big) + \frac{1}{2}\big\{ \Xi\,\lrcorner\,(R\wedge R), A^a \big\} \right) \ . \nonumber
    \end{align}
    Using \eqref{eq:D_Xi},\eqref{eq:omega_Xi}, these combine as:
    \begin{align}
    	\delta_\Xi L = D\left(\Xi\,\lrcorner\left(\frac{1}{2}\tilde\Psi R\wedge R + \tilde\phi_a R\wedge F^a\right)\right) + \frac{1}{2}\left\{ \Xi\,\lrcorner\left(\tilde\phi_a R\wedge R \right), A^a \right\} \approx 0 \ .
    \end{align}
	Finally, consider the ``colored diffeomorphisms'' \eqref{eq:xi_transformations}. From \eqref{eq:xi_transformations},\eqref{eq:delta_R_F}, we see that the field strengths transform as:
    \begin{align}
  	  \delta_\xi R = 0 \ ; \quad \delta_\xi\tilde\Psi = \xi_a\,\lrcorner\,D\tilde\phi^a \ ; \quad \delta_\xi F^a = D(\xi^a\,\lrcorner\,R) \ ; \quad \delta_\xi\tilde\phi^a = 0 \ .
    \end{align}
    Using this and \eqref{eq:D_R_F},\eqref{eq:anti_derivations}, we find that the two terms in $L$ transform as:
    \begin{align}
      \begin{split}
        \frac{1}{2}\delta_\xi\left(\tilde\Psi R\wedge R\right) &= \frac{1}{2}(\xi_a\,\lrcorner\,D\tilde\phi^a)R\wedge R  \ ; \\
        \delta_\xi\left(\tilde\phi_a R\wedge F^a\right) &= \frac{1}{2}\,\tilde\phi_a D\big(\xi^a\,\lrcorner\,(R\wedge R)\big) \ .
      \end{split}
    \end{align}
    Using \eqref{eq:D_Xi}, these variations combine into:
    \begin{align}
	  \delta_\xi L = \frac{1}{2} D\big(\xi^a\,\lrcorner\,(\tilde\phi_a R\wedge R)\big) \approx 0 \ .
    \end{align}
    This concludes our proof that all the gauge symmetries identified in \cite{Krasnov2021} survive the coupling of the GR-like and YM-like sectors in our Lagrangian \eqref{eq:L_integrand}, or, equivalently, \eqref{eq:HS_L}.
    
	\section{Lightcone gauge} \label{sec:lightconegauge}
	
	In this section, we construct a lightcone description of the covariant HS self-dual GR+YM theory \eqref{eq:HS_L}. This extends the result of \cite{Neiman2024} to include the YM-like sector. The logic of the construction is as follows. First, we focus on the right-handed degrees of freedom, encoded in the connections $\omega,A^a$. These are governed by self-contained (non-linear) equations of motion \eqref{eq:R_R_EOM}-\eqref{eq:R_F_EOM}, independent of the left-handed degrees of freedom $\Psi,\phi^a$. The main step is to define a lightcone ansatz for $\omega,A^a$, which reduces these fields with their many components (encoded in their dependence on the spinor $y^\alpha$) to a single scalar field for each helicity (which we'll encode as dependence on a \emph{scalar} parameter $u$ in place of $y^\alpha$). The consistency of this step will be verified by seeing that the field equations \eqref{eq:R_R_EOM}-\eqref{eq:R_F_EOM} also reduce to a single equation per helicity. In fact, these will be standard massless wave equations, plus some quadratic terms (which correspond to cubic interaction vertices). We then plug our ansatz for $\omega,A^a$ back into the covariant Lagrangian \eqref{eq:HS_L}, which includes the left-handed fields $\Psi,\phi^a$. This will have the effect of removing most of the Lagrangian's dependence on $\Psi,\phi^a$, except for one scalar component for each helicity, just like in the right-handed sector. The result is the desired lightcone Lagrangian.
	
    In the lightcone formulation, we will treat the dynamical fields as perturbations over a de Sitter background (specializing for concreteness to $\Lambda>0$). This background will be encoded as a particular (purely spin-2) value of the HS spin-connection $\omega$. We use Poincare coordinates $x^\mu$. In these coordinates, the de Sitter background metric is $\eta_{\mu\nu}/t^2$, where $\eta_{\mu\nu}$ is the Minkowski metric, and $t$ is the time component of $x^\mu$, normalized as $\eta^{\mu\nu}\partial_\mu t\partial_\nu t = -1$. For the AdS ($\Lambda<0$) case, one should replace $t$ with a spatial coordinate.
    
    We also introduce spinor indices for the Poincare coordinates, as $x^{\alpha\dot\alpha} \equiv \sigma^{\alpha\dot\alpha}_\mu x^\mu$, where $\sigma_\mu^{\alpha\dot\alpha}$ are the Pauli matrices for Minkowski space, satisfying $\sigma_\mu^{\alpha\dot\alpha}\sigma_{\nu\alpha\dot\alpha} = -2\eta_{\mu\nu}$. Here, the left-handed spinor indices $(\alpha,\beta,\dots)$ are raised/lowered with the flat spinor metric $\epsilon_{\alpha\beta}$ as in \eqref{eq:spinor_indices}, and likewise for the right-handed spinor indices $(\dot\alpha,\dot\beta,\dots)$. The spacetime index on $\sigma_\mu^{\alpha\dot\alpha}$ will be raised/lowered using the Minkowski metric $\eta_{\mu\nu}$. Finally, we'll use an arbitrary fixed spinor $q^\alpha$ to define the preferred lightlike direction of our lightcone gauge. 
    
    As mentioned above, we encode the (GR-like and YM-like) right-handed degrees of freedom in terms of generating functions $\Phi_+(x^\mu;u)$ and $\sigma_+^a(x^\mu;u)$, where the ``$+$'' signifies positive helicity, and the dependence on the scalar $u$ encodes the different spins:
    \begin{align}
	   \Phi_+(x;u) = \sum_s u^{s-1} \Phi_{+s}(x) \ ; \quad \sigma_+^a(x;u) = \sum_s u^{s-1} \sigma^a_{+s}(x) \ . \label{eq:Phi_sigma_decomposition}
    \end{align}
    In terms of these fields, we construct a lightcone ansatz for the GR-like connection $\omega$ and the YM-like connections $A^a$:
    \begin{align}
      \omega(x;y) &= -\frac{1}{2}dx^{\alpha\dot\alpha}\left(-y_\alpha y^\beta\partial_{\beta\dot\alpha}\ln t + q_\alpha q^\beta\partial_{\beta\dot\alpha}\Phi_+\!\left(x; u = \frac{\langle qy\rangle^2}{t} \right) \right) \ ; \label{eq:ansatz_omega} \\
      A^a(x;y) &= -\frac{1}{2}dx^{\alpha\dot\alpha}\,q_\alpha q^\beta\partial_{\beta\dot\alpha}\sigma_+^a\!\left(x; u = \frac{\langle qy\rangle^2}{t} \right) \ , \label{eq:ansatz_A}
    \end{align}
    where $\partial_{\alpha\dot\alpha}  = \sigma^\mu_{\alpha\dot\alpha}\partial_\mu$ is a spacetime gradient, and we defined the shorthand $\langle qy\rangle \equiv q_\alpha y^\alpha$. The first term in \eqref{eq:ansatz_omega} is the background connection mentioned above, i.e. the left-handed spin-connection of pure de Sitter space. The rest of \eqref{eq:ansatz_omega}-\eqref{eq:ansatz_A} is linear in the deformation parameters $\Phi_+(x;u)$ and $\sigma_+^a(x;u)$. We take the gradients in these deformation terms to act only on the $x^\mu$ argument of $\Phi_+(x;u)$ and $\sigma^a_+(x;u)$; in other words, we set $u = \langle qy\rangle^2/t$ only after taking the gradients.
    
    The curvature 2-form \eqref{eq:HS_R} of the GR-like connection \eqref{eq:ansatz_omega} reads:
    \begin{align}
       \begin{split}
 	    	R(x;y) &= \frac{1}{8}dx^{\alpha\dot\alpha}\wedge dx^{\beta\dot\beta}\big(\epsilon_{\dot\alpha\dot\beta}R_{\alpha\beta}(x;y) + \epsilon_{\alpha\beta}R_{\dot\alpha\dot\beta}(x;y) \big) \ ; \\
	    	R_{\alpha\beta}(x;y) &= \frac{2y_\alpha y_\beta}{t^2} - q_\alpha q_\beta\Box\Phi_+ 
	    	  + \frac{2q_{(\alpha} y_{\beta)} q^\gamma q^\delta}{\langle qy\rangle}\,\partial_\gamma{}^{\dot\beta}\ln t\,\partial_{\delta\dot\beta}(u\partial_u\Phi_+) \ ; \\
	    	R_{\dot\alpha\dot\beta}(x;y) &= q_\alpha q_\beta\big( {-\partial^\alpha{}_{\dot\alpha}\partial^\beta{}_{\dot\beta}\Phi_+} + 2\partial^\alpha{}_{(\dot\alpha}\ln t\,\partial^\beta{}_{\dot\beta)}(u\partial_u\Phi_+) \big) \ . 
       \end{split} \label{eq:R_ansatz}
    \end{align}
    Here, $\Box$ is the flat d'Alembertian $\Box\equiv \eta^{\mu\nu}\partial_\mu\partial_\nu$; the operator $u\partial_u\equiv u(\partial/\partial u)$ acts on each spin-$s$ component of $\Phi_+$ as multiplication by $s-1$; and, as before, we substitute $u=\langle qy\rangle^2/t$ \emph{after} taking all the derivatives. Similarly, for the curvature 2-form \eqref{eq:HS_F} of the YM-like connection \eqref{eq:ansatz_A}, we get:
    \begin{align}
		    F^a(x;y) &= \frac{1}{8}dx^{\alpha\dot\alpha}\wedge dx^{\beta\dot\beta}\big(\epsilon_{\dot\alpha\dot\beta}F^a_{\alpha\beta}(x;y) + \epsilon_{\alpha\beta}F^a_{\dot\alpha\dot\beta}(x;y) \big) \ ; \nonumber \\
	    	F^a_{\alpha\beta}(x;y) &= q_\alpha q_\beta\left(-\Box\sigma_+^a + \frac{1}{2}f^a{}_{bc}\,q^\gamma q^\delta \partial_\gamma{}^{\dot\alpha}\sigma_+^b\,\partial_{\delta\dot\alpha}\sigma_+^c \right)
	        	+ \frac{2q_{(\alpha} y_{\beta)} q^\gamma q^\delta}{\langle qy\rangle}\,\partial_\gamma{}^{\dot\beta}\ln t\,\partial_{\delta\dot\beta}(u\partial_u\sigma_+^a) \ ; \nonumber \\
	    	F^a_{\dot\alpha\dot\beta}(x;y) &= q_\alpha q_\beta\big( {-\partial^\alpha{}_{\dot\alpha}\partial^\beta{}_{\dot\beta}\sigma^a_+} + 2\partial^\alpha{}_{(\dot\alpha}\ln t\,\partial^\beta{}_{\dot\beta)}(u\partial_u\sigma^a_+) \big) \ . \label{eq:F_ansatz}
   \end{align}
   Note that $R$ in \eqref{eq:R_ansatz} depends on the GR-like deformation $\Phi_+$ only linearly, while $F^a$ in \eqref{eq:F_ansatz} doesn't depend on it at all. This happens because the would-be $\sim\Phi_+\Phi_+$ and $\sim\Phi_+\sigma_+^a$ contributions to the Poisson brackets $\{\omega,\omega\},\{\omega,A^a\}$ in \eqref{eq:HS_R}-\eqref{eq:HS_F} take the form $\sim q_\alpha q^\alpha$, which vanishes.

   Let us now plug \eqref{eq:R_ansatz}-\eqref{eq:F_ansatz} into the LHS of the field equations \eqref{eq:R_R_EOM}-\eqref{eq:R_F_EOM}. In Poincare coordinates, we can write the 4-forms $R\wedge R$ and $R\wedge F^a$ as: 
   \begin{align}
	  R\wedge R = \frac{i}{4}(R^{\dot\alpha\dot\beta}R_{\dot\alpha\dot\beta} - R^{\alpha\beta}R_{\alpha\beta})d^4x \ ; \quad R\wedge F^a = \frac{i}{4}(R^{\dot\alpha\dot\beta}F^a_{\dot\alpha\dot\beta} - R^{\alpha\beta}F^a_{\alpha\beta})d^4x \ ,
   \end{align}
   where $d^4x$ is the coordinate volume 4-form:
   \begin{align}
	  d^4x \equiv \frac{1}{24}\epsilon_{\mu\nu\rho\sigma}dx^\mu\wedge dx^\nu\wedge dx^\rho\wedge dx^\sigma \ .
   \end{align}
   From \eqref{eq:R_ansatz}-\eqref{eq:F_ansatz}, we read off the combinations $R^{\dot\alpha\dot\beta}R_{\dot\alpha\dot\beta} - R^{\alpha\beta}R_{\alpha\beta}$ and $R^{\dot\alpha\dot\beta}F^a_{\dot\alpha\dot\beta} - R^{\alpha\beta}F^a_{\alpha\beta}$ as:
   \begin{align}
   	  \begin{split}
  	    R^{\dot\alpha\dot\beta}R_{\dot\alpha\dot\beta} - R^{\alpha\beta}R_{\alpha\beta} ={}& \frac{4u}{t}\,\Box\Phi_+ \\
	      &+ q^\alpha q^\beta q^\gamma q^\delta\,\partial_{\alpha\dot\alpha}\partial_{\beta\dot\beta}\Phi_+ \big( \partial_\gamma{}^{\dot\alpha}\partial_\delta{}^{\dot\beta}\Phi_+ 
  	       - 4\,\partial_\gamma{}^{\dot\alpha}\ln t\,\partial_\delta{}^{\dot\beta}(u\partial_u\Phi_+) \big) \ ;
  	 \end{split} \label{eq:R_R_ansatz} \\
  	 \begin{split}
   	   R^{\dot\alpha\dot\beta}F^a_{\dot\alpha\dot\beta} - R^{\alpha\beta}F^a_{\alpha\beta} ={}& \frac{2u}{t}\left(\Box\sigma_+^a - \frac{1}{2}f^a{}_{bc}\,q^\alpha q^\beta\partial_\alpha{}^{\dot\alpha}\sigma_+^b\,\partial_{\beta\dot\alpha}\sigma_+^c\, \right) \\
         &+ q^\alpha q^\beta q^\gamma q^\delta\Big(\partial_{\alpha\dot\alpha}\partial_{\beta\dot\beta}\Phi_+ \big( \partial_\gamma{}^{\dot\alpha}\partial_\delta{}^{\dot\beta}\sigma_+^a 
         - 2\,\partial_\gamma{}^{\dot\alpha}\ln t\,\partial_\delta{}^{\dot\beta}(u\partial_u\sigma_+^a) \big) \\
         &\qquad\qquad\qquad - 2\,\partial_{\alpha\dot\alpha}\partial_{\beta\dot\beta}\sigma_+^a\,\partial_\gamma{}^{\dot\alpha}\ln t\,\partial_\delta{}^{\dot\beta}(u\partial_u\Phi_+) \Big) \ ,
     \end{split} \label{eq:R_F_ansatz}
   \end{align}
   where the explicit factors of $u$ are subject to the substitution $u = \langle qy\rangle^2/t$, just like the implicit ones inside $\Phi_+(x;u)$ and $\sigma_+^a(x;u)$. Crucially, all the $y^\alpha$-dependence in \eqref{eq:R_R_ansatz}-\eqref{eq:R_F_ansatz} is packaged as dependence on the scalar $u$. This allows us to write the field equations \eqref{eq:R_R_EOM}-\eqref{eq:R_F_EOM} as equations on $\Phi_+(x;u)$ and $\sigma_+^a(x;u)$:
   \begin{align}
	 \Box\Phi_+ ={}& \frac{q^\alpha q^\beta q^\gamma q^\delta}{u}\,\partial_{\alpha\dot\alpha}\partial_{\beta\dot\beta}\Phi_+\left(-\frac{t}{4}\,\partial_\gamma{}^{\dot\alpha}\partial_\delta{}^{\dot\beta}\Phi_+ 
    	  + \partial_\gamma{}^{\dot\alpha}t\,\partial_\delta{}^{\dot\beta}(u\partial_u\Phi_+) \right) \ ; \label{eq:Phi_equation} \\
     \begin{split}
       \Box\sigma^a_+ ={}& \frac{1}{2}f^a{}_{bc}\,q^\alpha q^\beta\partial_\alpha{}^{\dot\alpha}\sigma_+^b\,\partial_{\beta\dot\alpha}\sigma_+^c \\
           &+ \frac{q^\alpha q^\beta q^\gamma q^\delta}{u}\left(\partial_{\alpha\dot\alpha}\partial_{\beta\dot\beta}\Phi_+\left(-\frac{t}{2}\,\partial_\gamma{}^{\dot\alpha}\partial_\delta{}^{\dot\beta}\sigma^a_+ 
            + \partial_\gamma{}^{\dot\alpha}t\,\partial_\delta{}^{\dot\beta}(u\partial_u\sigma_+^a) \right) \right. \\
          &\qquad\qquad\qquad \left. \vphantom{\frac{t}{2}} + \partial_{\alpha\dot\alpha}\partial_{\beta\dot\beta}\sigma_+^a\,\partial_\gamma{}^{\dot\alpha}t\,\partial_\delta{}^{\dot\beta}(u\partial_u\Phi_+) \right) \ .
     \end{split} \label{eq:sigma_equation}
  \end{align}
  This is one scalar field equation per spin, which demonstrates the consistency of our ansatz \eqref{eq:ansatz_omega}-\eqref{eq:ansatz_A}.
  
  Having worked out the right-handed equations of motion, we're ready to re-introduce the linearized left-handed degrees of freedom $\Psi(x;y)$ and $\phi^a(x;y)$. Within the lightcone ansatz, these will enter the Lagrangian only through a single scalar component per spin. We define these negative-helicity components as:
  \begin{align}
  	\Phi_{-s}(x) \equiv -\frac{q^{\alpha_1}\dots q^{\alpha_{2s}}}{t^{s+1}}\,\Psi_{\alpha_1\dots\alpha_{2s}}(x) \ ; \quad \sigma^a_{-s}(x) \equiv -\frac{q^{\alpha_1}\dots q^{\alpha_{2s}}}{t^{s+1}}\,\phi^a_{\alpha_1\dots\alpha_{2s}}(x) \ . \label{eq:negative_helicities}
  \end{align}
  Plugging \eqref{eq:R_R_ansatz}-\eqref{eq:R_F_ansatz} back into the Lagrangian \eqref{eq:HS_L} and separating spin-by-spin, we finally obtain the lightcone Lagrangian as:
  \begin{align}
		S ={}& \int d^4x \sum_s\left(\Phi_{-s}\Box\Phi_{+s} + \sigma^a_{-s}\Box\sigma^a_{+s}\right) \nonumber \\
		&+ \frac{1}{2}f_{abc}\,q^\alpha q^\beta \!\! \sum_{s_2+s_3 = s_1+1} \sigma^a_{-s_1} \partial_{\alpha\dot\alpha}\sigma^b_{+s_2} \partial_\beta{}^{\dot\alpha}\sigma^c_{+s_3} \label{eq:S_lightcone} \\
		&+ q^\alpha q^\beta q^\gamma q^\delta \!\! \sum_{s_2+s_3 = s_1+2}\left(\Phi_{-s_1} \partial_{\alpha\dot\alpha}\partial_{\beta\dot\beta}\Phi_{+s_2}
		 \left(\frac{t}{4}\,\partial_\gamma{}^{\dot\alpha}\partial_\delta{}^{\dot\beta}\Phi_{+s_3} - (s_3-1)\partial_\gamma{}^{\dot\alpha}t\,\partial_\delta{}^{\dot\beta}\Phi_{+s_3} \right)\right. \nonumber \\
		&\left.\vphantom{\frac{t}{4}} \qquad\qquad\qquad\qquad\qquad + \sigma^a_{-s_1}\partial_{\alpha\dot\alpha}\partial_{\beta\dot\beta}\Phi_{+s_2}
          \left(\frac{t}{2}\,\partial_\gamma{}^{\dot\alpha}\partial_\delta{}^{\dot\beta}\sigma^a_{+s_3} - (s_3-1)\partial_\gamma{}^{\dot\alpha}t\,\partial_\delta{}^{\dot\beta}\sigma^a_{+s_3} \right) \right. \nonumber \\
		&\left.\vphantom{\frac{t}{4}} \qquad\qquad\qquad\qquad\qquad - (s_3-1)\sigma^a_{-s_1} \partial_{\alpha\dot\alpha}\partial_{\beta\dot\beta}\sigma^a_{+s_2}\partial_\gamma{}^{\dot\alpha}t\,\partial_\delta{}^{\dot\beta}\Phi_{+s_3} \right) \ . \nonumber
  \end{align}
  Here, the first line is the kinetic terms of the GR-like and YM-like fields; the second line is the YM-like cubic vertices, which only involve YM-like fields; the remaining lines are the GR-like cubic vertices, which involve both GR-like and YM-like fields. The sums over spins don't make explicit our restriction to even spins for GR-like fields and odd spins for YM-like fields, but this shouldn't introduce any confusion. Note that the sum of helicities in the kinetic terms, YM-like interactions and GR-like interactions is respectively $0$, $+1$ and $+2$; this is reflected in the powers of the auxiliary spinor $q^\alpha$. Note that the cubic action \eqref{eq:S_lightcone} is exact: there aren't any neglected higher-order vertices.
  
  The variations of the lightcone action \eqref{eq:S_lightcone} yield the following field equations, all with the massless wave operator $\Box$ on the LHS, and quadratic interaction terms on the RHS:
  \begin{itemize}
  	\item Varying w.r.t. $\Phi_{-s}$ yields the field equation \eqref{eq:Phi_equation} for $\Phi_{+s}$. In it, $\Box\Phi_{+s}$ receives GR-like contributions of the form $\sim\Phi_{+s_1}\Phi_{+s_2}$ with $s_1+s_2=s+2$.
  	\item Varying w.r.t. $\sigma^a_{-s}$ yields the field equation \eqref{eq:sigma_equation} for $\sigma^a_{+s}$. In it, $\Box\sigma^a_{+s}$ receives a YM-like contribution of the form $\sim f^a{}_{bc}\,\sigma^b_{+s_1}\sigma^c_{+s_2}$ with $s_1+s_2=s+1$, and GR-like contributions of the form $\sim \Psi_{+s_1}\sigma_{+s_2}$ with $s_1+s_2=s+2$.
  	\item Varying w.r.t. $\Phi_{+s}$ yields a field equation for $\Phi_{-s}$. In it, $\Box\Phi_{-s}$ receives GR-like contributions of the form $\sim\Phi_{-s_1}\Phi_{+s_2}$ and $\sim\sigma^a_{-s_1}\sigma^a_{+s_2}$ with $s_2-s_1=s+2$. Note that the last contribution, which mixes the GR-like and YM-like sectors, mirrors the sector-mixing gauge variations of $\Psi$ in \eqref{eq:theta_GR},\eqref{eq:xi_GR}.
  	\item Varying w.r.t. $\sigma^a_{+s}$ yields the field equation \eqref{eq:sigma_equation} for $\sigma^a_{-s}$. In it, $\Box\sigma^a_{-s}$ receives a YM-like contribution of the form $\sim f^a{}_{bc}\,\sigma^b_{-s_1}\sigma^c_{+s_2}$ with $s_2-s_1=s+1$, and GR-like contributions of the form $\sim \sigma^a_{-s_1}\Psi_{+s_2}$ with $s_2-s_1=s+2$.
  \end{itemize}
  Note that not all fields enter into each other's field equations. In fact, we can observe the same field-on-background hierachies that we described at the covariant level in section \ref{sec:covTh:hierarchy}. In particular, the field equation for $\Phi_{+2}$ receives non-linear contributions only from $\Phi_{+2}$ itself, and is just the lightcone field equation of self-dual GR from \cite{Neiman:2023bkq}. When this equation is satisfied, the spin-2 component of the ansatz \eqref{eq:ansatz_omega} becomes the left-handed spin connection of an Einstein spacetime with cosmological constant and purely right-handed Weyl curvature. The vielbein of this spacetime reads \cite{Neiman:2023bkq}:
  \begin{align}
 	e_\mu^{\alpha\dot\alpha} = \frac{1}{t}\sigma_\mu^{\alpha\dot\alpha} + \frac{1}{2}\sigma_\mu^{\beta\dot\beta}q^\alpha q_\beta q^\gamma q^\delta \left(\partial_\gamma{}^{\dot\alpha}\partial_{\delta\dot\beta}\Phi_{+2} 
 	  - \frac{2}{t}\,\partial_\gamma{}^{\dot\alpha}t\,\partial_{\delta\dot\beta}\Phi_{+2}\right) \ , \label{eq:e}
  \end{align}
  with inverse:
  \begin{align}
     e^\mu_{\alpha\dot\alpha} = t\sigma^\mu_{\alpha\dot\alpha} - \frac{t^2}{2}\sigma^\mu_{\beta\dot\beta}\,q_\alpha q^\beta q^\gamma q^\delta \left(\partial_\gamma{}^{\dot\beta}\partial_{\delta\dot\alpha}\Phi_{+2} 
       - \frac{2}{t}\,\partial_\gamma{}^{\dot\beta}t\,\partial_{\delta\dot\alpha}\Phi_{+2}\right) \ . \label{eq:e_inverse}
  \end{align}
  The first term in \eqref{eq:e}-\eqref{eq:e_inverse} describe pure de Sitter space, while the second term describes the self-dual deformation. Just like in \eqref{eq:ansatz_omega},\eqref{eq:R_ansatz}, the deformation terms are linear in the deformation parameter $\Phi_{+2}(x)$, despite the overall non-linearity of the theory.
  
  \subsection{Solving the field equation of higher-spin self-dual YM} \label{sec:covTh:HS_SD_YM}
  
  As an aside, let us consider the field equation \eqref{eq:sigma_equation} in the case of vanishing GR-like deformation $\Phi_+ = 0$. This gives the field equation for higher-spin self-dual YM on a pure de Sitter background:
  \begin{align}
  	\Box\sigma^a_+ = \frac{1}{2}f^a{}_{bc}\,q^\alpha q^\beta\partial_\alpha{}^{\dot\alpha}\sigma_+^b\,\partial_{\beta\dot\alpha}\sigma_+^c \ . \label{eq:HS_SD_YM_raw}
  \end{align}
  This is precisely the field equation of (standard, spin-1) self-dual YM. In particular, the warping factor $t$ of the de Sitter metric never appears: eq. \eqref{eq:HS_SD_YM_raw} might as well be an equation on flat spacetime. The only effect of higher spins is that $\sigma^a_+$ depends not only on spacetime position $x^\mu$, but also on the auxiliary scalar $u$ -- recall \eqref{eq:Phi_sigma_decomposition}. However, $u$ doesn't enter the field equation \eqref{eq:HS_SD_YM_raw} at all. This means we can solve \eqref{eq:HS_SD_YM_raw} as if we're in standard self-dual YM, and simply add in the $u$-dependence at the end. The perturbative solution of \eqref{eq:HS_SD_YM_raw} to all orders is well-known \cite{Bardeen1996,Rosly1997}. Let us recall it, following \cite{Albrychiewicz2021}. 
  
  We start by rewriting \eqref{eq:HS_SD_YM} without color indices:
  \begin{align}
	\Box\sigma_+ = q^\alpha q^\beta\partial_\alpha{}^{\dot\alpha}\sigma_+\,\partial_{\beta\dot\alpha}\sigma_+ \ , \label{eq:HS_SD_YM}
  \end{align}
  where $\sigma_+$ is now thought of as a matrix in (the defining representation of) the YM gauge algebra. Now, consider the linearized equation $\Box\sigma_+ = 0$. This will be solved in terms of lightlike 4-momenta $k^\mu=(k^t,\mathbf{k})$ with $k^t=\pm|\mathbf{k}|$, which in spinor indices take the form:
  \begin{align}
  	k_{\alpha\dot\alpha} = \lambda_\alpha\tilde\lambda_{\dot\alpha} \ ; \quad \tilde\lambda_{\dot\alpha} = \operatorname{sign}(k^t)\bar\lambda_{\dot\alpha} \ . \label{eq:spinor_helicity_4d}
  \end{align}
  Integrating over lightlike waves $e^{ik_\mu x^\mu} \equiv e^{ik\cdot x}$ with these 4-momenta with mode coefficients $c_+(\lambda^\alpha,\tilde\lambda^{\dot\alpha})$, we construct the general solution to $\Box\sigma^a_+ = 0$ as:
  \begin{align}
  	\sigma^{\text{lin}}_+(x) = \int_{k^2 = 0} \frac{d^3\mathbf{k}}{2k^t}\,\frac{c_+(\lambda,\tilde\lambda)}{\langle q\lambda\rangle^2}\,e^{ik\cdot x} \ , \label{eq:sigma_linear}
  \end{align}
  where the integration range is understood to include both positive-frequency and negative-frequency modes:
  \begin{align}
 	\int_{k^2 = 0} \equiv \int_{k^t=|\mathbf{k}|} + \int_{k^t=-|\mathbf{k}|} \ .
  \end{align}
  Now, the statement of \cite{Bardeen1996,Rosly1997} is that the non-linear equation \eqref{eq:HS_SD_YM} can be solved by augmenting the linearized solution \eqref{eq:sigma_linear} with non-linear perturbative corrections at all orders $n$, as:
  \begin{align}
  	\begin{split}
    	\sigma_+(x) = -\sum_{n=1}^\infty \prod_{m = 1}^n \left(\int_{k_m^2 = 0} \frac{d^3\mathbf{k}_m}{2k_m^t}\,c_+(\lambda_m,\tilde\lambda_m)\,e^{ik_m\cdot x}\right)
    	  \frac{1}{\langle q\lambda_1\rangle\langle\lambda_1\lambda_2\rangle\dots\langle\lambda_{n-1}\lambda_n\rangle\langle\lambda_n q\rangle} \ .
   	\end{split}
  \end{align}
  Higher spins can now be trivially incorporated by introducing $u$-dependence, as:
  \begin{align}
		\sigma_+(x;u) = -\sum_{n=1}^\infty \prod_{m = 1}^n \left(\int_{k_m^2 = 0} \frac{d^3\mathbf{k}_m}{2k_m^t}\,c_+(\lambda_m,\tilde\lambda_m;u)\,e^{ik_m\cdot x}\right)
		\frac{1}{\langle q\lambda_1\rangle\langle\lambda_1\lambda_2\rangle\dots\langle\lambda_{n-1}\lambda_n\rangle\langle\lambda_n q\rangle} \ . \label{eq:sigma_solution}
  \end{align}
  Of course, the full field equations from the lightcone action \eqref{eq:S_lightcone}, including the GR-like sector, can also be solved perturbatively. However, due to the factors of $t$ and $\partial_\mu t$ in the GR-like interactions, a closed-form solution such as \eqref{eq:sigma_solution} seems much more difficult to find.
	
  \section{Comparing lightcone gauges with a shared lightray} \label{sec:comparing_gauges}

  In this section, we consider how different lightcone gauges of the form described in section \ref{sec:lightconegauge} can be related to each other. This question naturally arises when we compute scattering amplitudes in the de Sitter static patch \cite{Albrychiewicz2021,Neiman:2023bkq}. Here, we'll describe and solve it as generally as we can, without direct reference to the static-patch problem. The more specific static-patch setup will come up in section \ref{sec:comparing_gauges:gaugetransf}.

  \subsection{Geometric picture and problem statement} \label{sec:comparing_gauges:geometry}
  
  As discussed in section \ref{sec:covTh:hierarchy}, the higher-spin theory \eqref{eq:HS_L} can be separated into the $s=2$ spin connection $\omega_{\alpha\beta}(x)$, and all other fields, namely $A^a,\Psi,\phi^a$ and the higher-spin components of $\omega$. In this separation, the spin connection $\omega_{\alpha\beta}$ is governed by standard self-dual GR, while the other fields are tensors living on this geometric background. In the lightcone gauge of section \ref{sec:lightconegauge}, the vielbein of the geometric background is given explicitly by \eqref{eq:e}-\eqref{eq:e_inverse}. Alternatively, we can set pure de Sitter space -- the first term in \eqref{eq:ansatz_omega} -- as background, and treat the spin-2 deformation term in \eqref{eq:ansatz_omega} as yet another tensor propagating on top of it. The background vielbein is then given by just the first term in \eqref{eq:e}-\eqref{eq:e_inverse}. Our statements in this section will hold equally well in both pictures: as detailed in \cite{Neiman:2023bkq} for self-dual GR, all the relevant differences will involve contractions of the spinor $q^\alpha$ with itself, and thus vanish. For simplicity of notation, regardless of which picture we choose, we'll denote the background vielbein as $e_\mu^{\alpha\dot\alpha}$ with inverse $e^\mu_{\alpha\dot\alpha}$, and the \emph{non-background} piece of the connection $\omega$ as $\Omega$, i.e. $\omega \equiv \omega_{\text{background}} + \Omega$.
  
  With a background geometry in hand, we can now make precise our claim that the ansatz \eqref{eq:ansatz_omega}-\eqref{eq:ansatz_A} defines a \emph{lightcone gauge}. To do this, we must gradually back away from the Poincare coordinates of \eqref{eq:ansatz_omega}-\eqref{eq:ansatz_A}, towards more covariant statements. First, we rescale the spinor $q^\alpha$, which was spacetime-independent in the Poincare coordinates, into:
  \begin{align}
  	w^\alpha \equiv \frac{q^\alpha}{\sqrt{t}} \ .
  \end{align}
  We thus rewrite the ansatz \eqref{eq:ansatz_omega}-\eqref{eq:ansatz_A} for the right-handed fields $\Omega,A^a$ as:
  \begin{align}
	\Omega(x;y) &\equiv -\frac{1}{2}dx^\mu e_\mu^{\alpha\dot\alpha} \sum_s y^{\alpha_1}\dots y^{\alpha_{2s-2}}\,\hat\Omega_{\alpha_1\dots\alpha_{2s-2}\alpha\dot\alpha}(x) \ ; \label{eq:ansatz_indices_omega} \\
	A^a(x;y) &\equiv -\frac{1}{2}dx^\mu e_\mu^{\alpha\dot\alpha} \sum_s y^{\alpha_1}\dots y^{\alpha_{2s-2}}\,\hat A^a_{\alpha_1\dots\alpha_{2s-2}\alpha\dot\alpha}(x) \ ; \label{eq:ansatz_indices_A} \\
	\hat\Omega_{\alpha_1\dots\alpha_{2s-2}\alpha\dot\alpha}(x) &= t\,w_{\alpha_1}\dots w_{\alpha_{2s-2}}w_\alpha w^\beta e^\nu_{\beta\dot\alpha}\partial_\nu\Phi_{+s}(x) \ ; \label{eq:Omega_ansatz_covariantly} \\
	\hat A^a_{\alpha_1\dots\alpha_{2s-2}\alpha\dot\alpha}(x) &= t\,w_{\alpha_1}\dots w_{\alpha_{2s-2}}w_\alpha w^\beta e^\nu_{\beta\dot\alpha}\partial_\nu\sigma^a_{+s}(x) \ , \label{eq:A_ansatz_covariantly}
  \end{align}    
  where we now take care to use spinor indices only for the internal flat tangent space. Similarly, we rewrite the left-handed components \eqref{eq:negative_helicities} as:
  \begin{align}
    \Phi_{-s}(x) \equiv -\frac{w^{\alpha_1}\dots w^{\alpha_{2s}}}{t}\,\Psi_{\alpha_1\dots\alpha_{2s}}(x) \ ; \quad \sigma^a_{-s}(x) \equiv -\frac{w^{\alpha_1}\dots w^{\alpha_{2s}}}{t}\,\phi^a_{\alpha_1\dots\alpha_{2s}}(x) \ . \label{eq:negative_helicities_covariantly}
  \end{align}
  The ansatz \eqref{eq:Omega_ansatz_covariantly}-\eqref{eq:A_ansatz_covariantly} is adapted to a foliation of spacetime into lightlike hypersurfaces w.r.t. the background vielbein $e_\mu^{\alpha\dot\alpha}$. In the Poincare coordinates, these hypersurfaces appear as flat hyperplanes $\sigma_\mu^{\alpha\dot\alpha} q_\alpha \bar q_{\dot\alpha} x^\mu = \text{const}$. However, in the actual geometry defined by $e_\mu^{\alpha\dot\alpha}$, they are the \emph{lightcones} of ordinary points inside the spacetime (which, in the Poincare coordinates, appear as points at lightlike infinity; see e.g. \cite{Neiman:2023bkq}). Along each lightray within one of these lightcones, we may consider the coordinate $t$ as a null time (even though the full gradient $\partial_\mu t$ is timelike). The rate of this null time is fixed geometrically, up to a constant, by the fact that its inverse $1/t$ is an \emph{affine} null time, with $1/t = 0$ marking the lightcone's origin point \cite{Neiman:2023bkq}. The tangent vector to the lightcone's lightrays, whose scaling is fixed w.r.t. $t$ as $\ell^\mu\partial_\mu t = \text{const}$, reads:
  \begin{align}
  	\ell^\mu \equiv -e^\mu_{\alpha\dot\alpha}w^\alpha \bar w^{\dot\alpha} \ . \label{eq:ell}
  \end{align}
  In the Poincare coordinates, this is just the constant vector $\ell^\mu = -\sigma^\mu_{\alpha\dot\alpha}q^\alpha\bar q^\alpha$. We can further normalize $\ell^\mu\partial_\mu t$ to unity, by fixing:
  \begin{align}
  	\sigma^\mu_{\alpha\dot\alpha}q^\alpha\bar q^{\dot\alpha}\partial_\mu t = -1 \ .
  \end{align}
  With these ingredients, we can state the lightcone gauge conditions geometrically, without referring to the Poincare coordinates. The conditions can be stated w.r.t. a single lightcone hypersurface, rather than the entire foliation. Moreover, we can consider just the components of $\hat\Omega_{\alpha_1\dots\alpha_{2s-2}\alpha\dot\alpha},\hat A^a_{\alpha_1\dots\alpha_{2s-2}\alpha\dot\alpha}$ that are tangential to the lightcone, i.e. along the $w^\alpha(\dots)^{\dot\alpha}$ or $(\dots)^\alpha\bar w^{\dot\alpha}$ directions. For these components, the ansatz \eqref{eq:Omega_ansatz_covariantly}-\eqref{eq:A_ansatz_covariantly} reads:
  \begin{align}
  	w^\alpha\hat\Omega_{\alpha_1\dots\alpha_{2s-2}\alpha\dot\alpha}(x) &= 0 \ ; \quad
  	\bar w^{\dot\alpha}\hat\Omega_{\alpha_1\dots\alpha_{2s-2}\alpha\dot\alpha}(x) = -t\,w_{\alpha_1}\dots w_{\alpha_{2s-2}}w_\alpha\,\ell^\nu\partial_\nu\Phi_{+s}(x) \ ; \label{eq:Omega_ansatz_lightcone} \\
  	w^\alpha\hat A^a_{\alpha_1\dots\alpha_{2s-2}\alpha\dot\alpha}(x) &= 0 \ ; \quad 
  	\bar w^{\dot\alpha}\hat A^a_{\alpha_1\dots\alpha_{2s-2}\alpha\dot\alpha}(x) = -t\,w_{\alpha_1}\dots w_{\alpha_{2s-2}}w_\alpha\,\ell^\nu\partial_\nu\sigma^a_{+s}(x) \ . \label{eq:A_ansatz_lightcone}
  \end{align}    
  The gauge conditions can now be stated as follows.
  
  \paragraph{Geometric statement of lighcone gauge conditions.} 
  Consider a lightcone within the background geometry defined by the vielbein $e_\mu^{\alpha\dot\alpha}$ (either pure de Sitter or its self-dual deformation). Along each lightray within this lightcone, define a null coordinate $t$ such that $1/t$ is affine, with $1/t=0$ being the lightcone's origin. Define a tangent null vector along the lightrays via $\ell^\mu\partial_\mu t = 1$. Define its spinor square root $w^\alpha$ via \eqref{eq:ell}. Fix the complex phase of $w^\alpha$ along each lightray by demanding that the covariant derivative $\ell^\mu\nabla_\mu w^\alpha$ is a \emph{real} multiple of $w^\alpha$. In terms of these quantities, lightcone gauge is defined by imposing the ansatz \eqref{eq:Omega_ansatz_lightcone}-\eqref{eq:A_ansatz_lightcone} throughout the lightcone. The individual lightcone fields with positive helicities $\Phi_{+s},\sigma^a_{+s}$ are then defined via \eqref{eq:Phi_sigma_decomposition}, while the negative-helicity fields $\Phi_{-s},\sigma^a_{-s}$ are defined via \eqref{eq:negative_helicities_covariantly}.
  \newline
  
  In fact, the lightcone gauge of section \ref{sec:lightconegauge} is slightly more specific than this, as it implies some relationships between the scaling of $t$ across different lightrays, and similarly for the phase of $w^\alpha$. However, this subtlety will not be important in the following. The use of actual lightcones of bulk points makes our class of lightcone gauges (which originated in \cite{Neiman:2023bkq}) more general than the usual ones in the literature, which use either null hyperplanes in Minkowski space, or the lightcones of boundary points in Anti-de Sitter (e.g. \cite{Metsaev:2018xip}). This greater generality enables us to ask a question which never arises in the more usual, constrained setup, and is relevant for the scattering problem in the de Sitter static patch.
  
  \paragraph{Problem statement.}
  Consider two lightcone gauges as above, defined w.r.t. to two different lightcones that share a lightray. On the shared lightray, what is the relationship between the lightcone fields $\Phi_{\pm s},\sigma^a_{\pm s}$ in these two gauges?
  \newline
  
  The relevant case for static-patch scattering \cite{Albrychiewicz2021,Neiman:2023bkq} is the limit where one of the two lightcones is a cosmological horizon, i.e. its origin point is at the spacetime's conformal boundary. In fact, solving this limiting case is equivalent to solving the more general problem above. This is because the general case can be obtained by first relating the fields between one lightcone and a cosmological horizon, and then again between the horizon and a second lightcone. We will work with the general case in section \eqref{sec:comparing_gauges:field_strengths}, and with the limiting case in section \ref{sec:comparing_gauges:gaugetransf}.
  
  \subsection{Solution via gauge-invariant field strengths} \label{sec:comparing_gauges:field_strengths}
  
  In \cite{Albrychiewicz2021,Neiman:2023bkq}, we solved the problem of comparing lightcone gauges by passing from gauge potentials to gauge-covariant field strengths (which are actually gauge-invariant at linearized order), namely the YM field strength and the GR Weyl tensor. In the present higher-spin context, this approach runs into a problem: we don't know how to define the analogous field strengths for the right-handed degrees of freedom encoded in the potentials $\Omega,A^a$. In particular, these desired field strengths are \emph{not} just the curvature 2-forms \eqref{eq:HS_R}-\eqref{eq:HS_F} (except in the $s=1$ YM case). Indeed, those carry $2s-2$ left-handed spinor indices at each spin $s$, and transform non-trivially under the gauge transformations \eqref{eq:Theta_GR}-\eqref{eq:xi_YM}, already at the linearized level. The desired field strengths should instead be generalizations of the right-handed Weyl tensor, with all right-handed indices, removed from the 2-forms \eqref{eq:HS_R}-\eqref{eq:HS_F} by a tower of $2s-2$ spacetime derivatives.
  
  Fortunately, we can use a workaround. First, if we stay at linearized order, the gauge-invariant field strengths are trivial to write down, and we can proceed as in \cite{Albrychiewicz2021,Neiman:2023bkq}. Second, we can use causality and angular momentum conservation, along with the simple helicity structure of our self-dual theory \eqref{eq:HS_L}, to argue that the linearized presecription does not receive non-linear corrections. 
  
  We will denote quantities in the ``second'' lightcone gauge with a tilde. Recalling that $1/t$ is an affine coordinate on the shared lightray, its rate of flow $d/d(1/t)$ must be the same in both gauges, up to a constant:
  \begin{align}
  	\frac{d(1/\tilde t)}{d(1/t)} = \frac{t^2}{\tilde t^2}\frac{d\tilde t}{dt} = \text{const} \ . \label{eq:affine_ratio}
  \end{align}
  We fix the phase of the spinor $w^\alpha$ on the shared ray to be the same in both gauges. Then the values of $\ell^\mu$ and its square root $w^\alpha$ in the two gauges are related as:
   \begin{align}
  	\tilde\ell^\mu = \frac{dt}{d\tilde t}\,\ell^\mu \ ; \quad \tilde w^\alpha = \sqrt{\frac{dt}{d\tilde t}}\,w^\alpha \ . \label{eq:w_relation}
  \end{align}
  Let us now focus on the linearized theory, working at linearized order in the deformations $\Phi_{\pm s},\sigma_{\pm s}$ (in particular, this means we take the background geometry to be pure de Sitter space). Since the GR-like and YM-like sectors are fully equivalent at the linearized level, we will focus on the GR-like sector for brevity. Let us consider first the left-handed fields $\Psi$, as these are already field strengths that are gauge-invariant at the linearized level:
  \begin{align}
	\tilde\Psi_{\alpha_1\dots\alpha_{2s}}(x) = \Psi_{\alpha_1\dots\alpha_{2s}}(x) \ .
  \end{align}
  Together with \eqref{eq:negative_helicities_covariantly},\eqref{eq:w_relation}, this implies a proportionality between the lightcone fields $\Phi_{-s}(x)$ on the shared ray in the two gauges:
  \begin{align}
  	\tilde\Phi_{-s} = \frac{t}{\tilde t}\left(\frac{dt}{d\tilde t}\right)^s\Phi_{-s} \ . \label{eq:Phi_-}
  \end{align}
  Let us now turn to the right-handed degrees of freedom. At the linearized level, it is straightforward to write the gauge-invariant right-handed field strengths corresponding to the lightcone ansatz \eqref{eq:Phi_sigma_decomposition}-\eqref{eq:ansatz_omega}. In flat spacetime (see e.g. \cite{Sharapov2022a}), one would simply apply $2s-1$ spacetime derivatives $\partial^{\alpha_1}{}_{\dot\alpha_1}\dots\partial^{\alpha_{2s-2}}{}_{\dot\alpha_{2s-2}}\partial^\alpha{}_{\dot\beta}$ to the potentials \eqref{eq:Omega_ansatz_covariantly}, contracting all the left-handed spinor indices and symmetrizing the right-handed ones. Since the linearized field strengths are conformal, we can just use the flat-spacetime prescription in the Poincare coordinates, and apply appropriate powers of the conformal factor $t$. The result reads (with sign factors chosen for convenience):
  \begin{align}
    	\Psi_{\dot\alpha_1\dots\dot\alpha_{2s}} = (-t)^{s+1}q^{\alpha_1}\dots q^{\alpha_{2s}}\partial_{\alpha_1\dot\alpha_1}\dots\partial_{\alpha_{2s}\dot\alpha_{2s}}\Phi_{+s} \ , \label{eq:RH_Psi}
    \end{align}
  where the spinor indices on the LHS are understood to live in the internal flat tangent space. Contracting with $\bar w^{\dot\alpha}$ analogously to \eqref{eq:negative_helicities_covariantly}, we get an expression that no longer requires Poincare coordinates, and can be written in the more covariant language of section \ref{sec:comparing_gauges:geometry}:
  \begin{align}
  	\bar w^{\dot\alpha_1}\dots\bar w^{\dot\alpha_{2s}}\Psi_{\dot\alpha_1\dots\dot\alpha_{2s}} = -t (\ell^\mu\partial_\mu)^s\,\Phi_{+s} = -t\left(\frac{d}{dt}\right)^{2s}\!\Phi_{+s} = -\frac{1}{t^{2s}}\left(\frac{d}{d(1/t)}\right)^{2s}\frac{\Phi_{+s}}{t^{2s-1}} \ . \label{eq:RH_Psi_components}
  \end{align}
  Here, in the second equality, we treat $t$ as a coordinate along the shared lightray, so that $\ell^\mu\partial_\mu$ becomes $d/dt$. In the last equality, we used a differential identity for functions $f(t)$ of one variable, which is easy to prove by induction:
  \begin{align}
  	\left(\frac{d}{dt}\right)^n\!f = \frac{(-1)^n}{t^{n+1}}\left(\frac{d}{d(1/t)}\right)^n\frac{f}{t^{n-1}} \ .
  \end{align}
  Now, the components \eqref{eq:RH_Psi_components} in the two gauges should again be proportional to each other via \eqref{eq:w_relation}, so that:
   \begin{align}
  	\frac{1}{\tilde t^{2s}}\left(\frac{d}{d(1/\tilde t)}\right)^{2s}\frac{\tilde\Phi_{+s}}{\tilde t^{2s-1}} = \left(\frac{dt}{d\tilde t}\right)^s\,\frac{1}{t^{2s}} \left(\frac{d}{d(1/t)}\right)^{2s}\frac{\Phi_{+s}}{t^{2s-1}} \ .
  \end{align}
  Using the constant ratio \eqref{eq:affine_ratio} between $d(1/t)$ and $d(1/\tilde t)$, this simplifies into:
   \begin{align}
	\left(\frac{d}{d(1/t)}\right)^{2s}\frac{\tilde\Phi_{+s}}{\tilde t^{2s-1}} = \left(\frac{d(1/\tilde t)}{d(1/t)}\right)^s \left(\frac{d}{d(1/t)}\right)^{2s}\frac{\Phi_{+s}}{t^{2s-1}} \ .
  \end{align}
  Again using $d(1/\tilde t)/d(1/t) = \text{const}$, this can be integrated trivially as:
   \begin{align}
	\frac{\tilde\Phi_{+s}}{\tilde t^{2s-1}} &= \left(\frac{d(1/\tilde t)}{d(1/t)}\right)^s \frac{\Phi_{+s}}{t^{2s-1}} \\ 
	\quad \Longrightarrow \quad \tilde\Phi_{+s} &= \frac{t}{\tilde t}\left(\frac{d\tilde t}{dt}\right)^s\Phi_{+s} \ . \label{eq:Phi_+}
   \end{align}
   Thus, despite the appearance of spacetime derivatives in the intermediate steps, we find that $\tilde\Phi_{+s}$ is simply proportional to $\Phi_{+s}$ \emph{at the same point}, just as with the left-handed fields \eqref{eq:Phi_-}. In fact, \eqref{eq:Phi_-} and \eqref{eq:Phi_+} combine nicely into a unified result (along with its YM-like analogue):
   \begin{align}
   	 \tilde\Phi_{\pm s} = \frac{t}{\tilde t}\left(\frac{d\tilde t}{dt}\right)^{\pm s}\Phi_{\pm s} \ ; \quad \tilde\sigma^a_{\pm s} = \frac{t}{\tilde t}\left(\frac{d\tilde t}{dt}\right)^{\pm s}\sigma^a_{\pm s} \ . \label{eq:linearized_result}
   \end{align}
   Our next claim is that the linearized result \eqref{eq:linearized_result} \emph{persists at the non-linear level without corrections}. This follows from an argument in several steps:
  \begin{enumerate}
  	\item The covariant theory \eqref{eq:HS_L} inherits the standard two-derivative structure from self-dual GR+YM. As a result, initial data on a lightcone determines the fields throughout the spacetime as usual. The lightcone fields $\Phi_{\pm s},\sigma^a_{\pm s}$ inherit this property from the covariant fields via the relations \eqref{eq:negative_helicities_covariantly},\eqref{eq:Omega_ansatz_lightcone}-\eqref{eq:A_ansatz_lightcone}. Therefore, the fields on the shared lightray in one gauge must be determined \emph{somehow} by the fields \emph{throughout the lightcone} in the other gauge.
  	\item Again due to the standard two-derivative structure, the covariant theory \eqref{eq:HS_L} is \emph{causal}. As a result, the fields on the shared lightray in one gauge must be determined by the fields \emph{on the same lightray} in the other gauge, since all the other lightrays on the lightcone are spacelike-separated from the shared one. This is again inherited by the lightcone fields $\Phi_{\pm s},\sigma^a_{\pm s}$, since the relations \eqref{eq:negative_helicities_covariantly},\eqref{eq:Omega_ansatz_lightcone}-\eqref{eq:A_ansatz_lightcone} involve derivatives only along the lightray.
  	\item Any non-linear corrections to \eqref{eq:linearized_result} must conserve angular momentum. But, since the corrections must be localized on the lightray, this angular momentum can't be orbital, i.e. it can't arise from spatial derivatives. Thus, the non-linear corrections must preserve \emph{helicity}.
  	\item The non-linear corrections must arise from the interactions in the theory. Therefore, any such corrections require \emph{interactions with a vanishing sum of helicities}.
  	\item However, the theory \eqref{eq:HS_L},\eqref{eq:S_lightcone} does not contain such interactions! Specifically, all the vertices in \eqref{eq:S_lightcone} have a \emph{positive} sum of helicities ($+1$ for YM-like, $+2$ for GR-like). Therefore, there can be no non-linear corrections to \eqref{eq:linearized_result}.
  \end{enumerate}
	
  \subsection{Exploring the gauge transformation directly} \label{sec:comparing_gauges:gaugetransf}

  In this subsection, we take a different approach to the gauge-comparison problem defined in section \ref{sec:comparing_gauges:geometry}, which will reproduce the result from section \ref{sec:comparing_gauges:field_strengths}. This alternative approach is incomplete, as it will include some assumptions we cannot justify. The idea is to explicitly construct the gauge transformation from one lightcone gauge to the other, by ensuring in a ``minimal'' way that the necessary gauge conditions hold, not on the entire lightcone, but on the shared lightray \emph{and on a lightray infinitesimally displaced from it, to first order in the displacement}. We manage to do this on the deformed background geometry \eqref{eq:e}, and without any linearization in the fields. However, we were unable to prove (aside from some simple cases, which we'll mention briefly) that this ``greedy'' procedure is in fact consistent with fixing the correct gauge on the entire lightcone.
  
  \subsubsection{Geometric setup}
  
  For simplicity, we focus on the GR-like sector. Thus, the only fields are the connection $\omega = \omega_{\text{background}} + \Omega$ and the 0-form $\Psi$, and the only gauge transformations are the HS Lorentz rotations \eqref{eq:Theta_GR} and the HS diffeomorphisms \eqref{eq:Xi_GR}. For the ``first'' of the two lightcone gauges, we use the explicit ansatz \eqref{eq:ansatz_omega}, and use its associated Poincare coordinates throughout. Without loss of generality, we focus on the specific lightcone hypersurface:
  \begin{align}
  	\sigma_\mu^{\alpha\dot\alpha}q_\alpha q_{\dot\alpha}x^\mu = 0 \ . \label{eq:lightcone}
  \end{align} 
  For the ``second'' lightcone gauge, we take the limit relevant for static patch scattering, namely we take its origin to lie at conformal infinity, so that the lightcone itself is a cosmological horizon. In pure de Sitter space, the location of this horizon can be taken as the lightcone $\eta_{\mu\nu}x^\mu x^\nu = 0$ of the boundary point $x^\mu = 0$ (recall that in Poincare coordinates, the conformal boundary is at $t=0$). Here, we do \emph{not} assume a pure de Sitter geometry, and we don't have a closed-form expression for the horizon's location. However, the \emph{shared lightray} between the horizon and the lightcone \eqref{eq:lightcone} can be fixed, without loss of generality, as:
  \begin{align}
   x_0^\mu \sim \sigma^\mu_{\alpha\dot\alpha}q^\alpha\bar q^{\dot\alpha} \ , \label{eq:shared_ray}
 \end{align}
 where the ``$0$'' subscript indicates that we're on the shared ray.
  
  In the ``first'' gauge, adapted to the lightcone \eqref{eq:lightcone}, the $t$ coordinate from section \ref{sec:comparing_gauges:geometry}, whose inverse is affine along the lightrays, is, by construction, the same as the Poincare coordinate $t$. In the ``second'' gauge, adapted to the horizon, we are effectively always ``zoomed in'' on a relatively tiny interval of affine time, since the horizon's origin on the conformal boundary is at infinite affine distance from us. In this ``zoomed in'' regime, the inverse of an affine coordinate can be approximated linearly as simply affine, shifted by a very large constant. Thus, as the horizon's analog of the $t$ coordinate from section \ref{sec:comparing_gauges:geometry}, we can use:
  \begin{align}
  	\tilde t = \text{const} - \frac{1}{t} \ , \label{eq:tilde_t}
  \end{align}
  with the understanding that the constant term is much larger than the $1/t$ term. The lightlike tangent to the shared lightray \eqref{eq:shared_ray} that's scaled w.r.t. $\partial_\mu\tilde t$ reads:
  \begin{align}
  	\tilde\ell^\mu_0 = -t^2\sigma^\mu_{\alpha\dot\alpha}q^\alpha\bar q^{\dot\alpha} \ . \label{eq:ell_0}
  \end{align}
  On the shared ray, the tangent space to both the lightcone and the horizon consists of the lightlike vector \eqref{eq:ell_0}, along with two (complex) spatial vectors:
  \begin{align}
	m^\mu = \frac{t}{2}\,\sigma^\mu_{\alpha\dot{\alpha}}q^\alpha \bar\chi^{\dot\alpha} \ ; \quad 
	\bar{m}^\mu = \frac{t}{2}\,\sigma^\mu_{\alpha\dot{\alpha}} \chi^\alpha\bar{q}^{\dot{\alpha}} \ , \label{eq:m}
  \end{align}
  where we introduced a second constant spinor $\chi^\alpha$ that's normalized w.r.t. $q^\alpha$ as $q_\alpha\chi^\alpha = 1$, and likewise for $\bar\chi^{\dot\alpha}$ and $\bar q^{\dot\alpha}$.
  
  In pure de Sitter space, the two basis vectors \eqref{eq:m} are both null, with the totally-null bivectors $\tilde\ell_0\wedge m$ and $\tilde\ell_0\wedge\bar m$ left-handed and right-handed respectively. Moreover, we can define totally-null planes along these bivectors, as $x^\mu = \sigma^\mu_{\alpha\dot\alpha}q^\alpha(\dots)^{\dot\alpha}$ and $x^\mu = \sigma^\mu_{\alpha\dot\alpha}(\dots)^\alpha\bar q^{\dot\alpha}$ respectively. Like the original ray \eqref{eq:shared_ray}, these totally-null planes are also shared between the lightcone \eqref{eq:lightcone} and the horizon. Under the right-handed deformation \eqref{eq:e} of the de Sitter geometry, all this remains true of $m^\mu$ and the left-handed plane $x^\mu = \sigma^\mu_{\alpha\dot\alpha}q^\alpha(\dots)^{\dot\alpha}$, which remains undeformed, but not of their right-handed analogues $\bar m^\mu$ and $x^\mu = \sigma^\mu_{\alpha\dot\alpha}(\dots)^\alpha\bar q^{\dot\alpha}$.
  
  Along the shared left-handed plane, the lightcone gauge w.r.t. the lightcone \eqref{eq:lightcone} also functions as a lightcone gauge w.r.t. the horizon, with no need for any gauge transformation. Thus, we will focus on finding a gauge transformation that satisfies the horizon's lightcone gauge conditions \emph{at points removed from the shared lightray infinitesimally along $\bar m^\mu$}. In pure de Sitter space, we can just focus on points of the form $x^\mu = x_0^\mu + \bar h m^\mu + h\bar m^\mu$, where $h,\bar h$ are infinitesimal constants; in the pure de Sitter geometry, these points form a lightray, neighboring to \eqref{eq:shared_ray}, along the horizon $\eta_{\mu\nu}x^\mu x^\nu = 0$. In the deformed geometry \eqref{eq:e}, this neighboring lightray will itself get deformed. However, if we assume that the ray is undeformed at some initial point (e.g. at inifinity, or at the static patch's bifurcation sphere as in \cite{Neiman:2023bkq})), then the deformed ray will be of the form: 
  \begin{align}
  	x^\mu = x_0^\mu + \bar h m^\mu + h\bar m^\mu + h\sigma^\mu_{\alpha\dot\alpha}q^\alpha(\dots)^{\dot\alpha} \ . \label{eq:next_x_raw}
  \end{align}
  This is because the deformation to the Christoffel symbols $\Gamma^\mu_{\nu\rho}$ from \eqref{eq:e} is always along the $\sigma^\mu_{\alpha\dot\alpha}q^\alpha(\dots)^{\dot\alpha}$ plane, in at least one of the indices $\mu,\nu,\rho$. The lightlike tangent to the horizon ray \eqref{eq:next_x_raw} reads:
  \begin{align}
  	\tilde\ell^\mu \equiv \frac{dx^\mu}{d\tilde t} = \tilde\ell^\mu_0 + \bar h tm^\mu + ht\bar m^\mu + h\sigma^\mu_{\alpha\dot\alpha}q^\alpha(\dots)^{\dot\alpha} \ . \label{eq:next_ell_raw}
  \end{align}
  The $h\sigma^\mu_{\alpha\dot\alpha}q^\alpha(\dots)^{\dot\alpha}$ term in \eqref{eq:next_x_raw} can be decomposed into a piece along $x_0^\mu$, and a piece along $m^\mu$. At any single point, the first of these can be absorbed into the definition of $x_0^\mu$, while the second can be cancelled against the $\bar h m^\mu$ term, by appropriately tuning $\bar h$ in proportion to $h$. Thus, even in the deformed geometry, we can simply focus on points removed from the shared ray along the $\bar m^\mu$ direction:
  \begin{align}
  	x^\mu = x_0^\mu + h\bar m^\mu \ , \label{eq:next_x}
  \end{align}
  while being mindful that the set of such points may not form a lightray. Instead, we read off from \eqref{eq:next_ell_raw} that the tangent to the horizon lightray passing through \eqref{eq:next_x} takes the form:
  \begin{align}
	\tilde\ell^\mu = \tilde\ell^\mu_0 + ht\bar m^\mu + h\sigma^\mu_{\alpha\dot\alpha}q^\alpha(\dots)^{\dot\alpha} \ , \label{eq:next_ell}
  \end{align}
  where we absorbed the $\bar h tm^\mu$ term into $h\sigma^\mu_{\alpha\dot\alpha}q^\alpha(\dots)^{\dot\alpha}$. From now on, we work to first order in small $h$. The spinor square root $\tilde w^\alpha$ of the lightlike tangent \eqref{eq:next_ell} reads:
  \begin{align}
	\tilde\ell^\mu = -e^\mu_{\alpha\dot\alpha}\tilde w^\alpha\bar{\tilde w}^{\dot\alpha} \ ; \quad 
	\tilde w^\alpha = \sqrt{t}\left((1 + O(h))q^\alpha - \frac{h}{2}\,\chi^\alpha \right) \ . \label{eq:next_w}
  \end{align}
  
  \subsubsection{Computing the gauge transformation}
  
  Let $\tilde\Omega_\mu(x;y)$ denote the gauge-transformed version of the 1-form $\Omega_\mu(x;y)$, i.e. its value in the gauge adapted to the horizon (while $\Omega_\mu(x;y)$ itself refers to the original gauge, adapted to the lightcone \eqref{eq:lightcone}). We write the gauge conditions \eqref{eq:Omega_ansatz_lightcone} in the horizon-adapted gauge as:
  \begin{align}
  	\tilde w^\alpha e^\mu_{\alpha\dot\alpha}\tilde\Omega_\mu(x;y) &= 0 \ ; \label{eq:general_conditions_spacetime} \\
  	\tilde w^\beta\partial_\beta\big(\bar{\tilde w}^{\dot\alpha} e^\mu_{\alpha\dot\alpha}\tilde\Omega_\mu(x;y)\big) &= 0 \ , \label{eq:general_conditions_spinor}
  \end{align}
  where we recall that $\partial_\alpha$ is a derivative w.r.t. the spinor $y^\alpha$. Note that \eqref{eq:general_conditions_spacetime}-\eqref{eq:general_conditions_spinor} capture all the restrictions on $\tilde\Omega_\mu$ implied by \eqref{eq:Omega_ansatz_lightcone}. The only additional content of \eqref{eq:Omega_ansatz_lightcone} is the definition of the lightcone fields $\tilde\Phi_{+s}$, which we will get to in section \ref{sec:comparing_gauges:gaugetransf:comparison}.
  
  On the shared left-handed plane, and in particular on the shared ray \eqref{eq:shared_ray}, the conditions \eqref{eq:general_conditions_spacetime}-\eqref{eq:general_conditions_spinor} already hold in the original gauge. We will thus choose the gauge-transformation parameters $\Theta(x;y),\Xi^\mu(x;y)$ on the shared left-handed plane to vanish. As an immediate consequence, the left-handed field $\Psi(x_0;y)$ on the shared ray remains invariant under the gauge transformation, since its transformation law doesn't involve gradients of $\Theta,\Xi^\mu$. We therefore focus purely on the 1-form $\Omega_\mu/\tilde\Omega_\mu$, whose gauge transformation \emph{does} involve a gradient $\partial_\mu\Theta$. Since we took the transformation parameters $\Theta,\Xi^\mu$ to vanish on the shared ray, their values at the neighboring points \eqref{eq:next_x} will be infinitesimal of order $O(h)$. This allows us to use the gauge transformations in their infinitesimal form \eqref{eq:Theta_GR},\eqref{eq:Xi_GR}. Out of the gauge conditions \eqref{eq:general_conditions_spacetime}-\eqref{eq:general_conditions_spinor}, we focus on enforcing the following components at the neighboring points \eqref{eq:next_x}:
  \begin{align}
	\tilde\Omega_{\tilde\ell}(x;y) &= 0 \ ; \label{eq:condition_spacetime} \\
	\tilde w^\alpha\partial_\alpha\tilde\Omega_{\bar m}(x;y) &= 0 \ , \label{eq:condition_spinor}
  \end{align}
  where we introduce the shorthand $\tilde\Omega_v \equiv v^\mu\tilde\Omega_\mu$ for contracting vectors with 1-forms. Note that on the defirmed geometry, the condition \eqref{eq:condition_spinor} is not a component of \eqref{eq:general_conditions_spinor}, but it \emph{is} a linear combination of \eqref{eq:general_conditions_spacetime} and \eqref{eq:general_conditions_spinor}. In our ``greedy'' gauge-transformation procedure, the components \eqref{eq:condition_spacetime}-\eqref{eq:condition_spinor} will be sufficient to determine the field component $\tilde\Omega_{\bar m}$ on the shared ray, which in turn will be sufficient for the desired comparison of the lightcone fields $\Phi_{+s}$ in the two gauges. Writing \eqref{eq:condition_spacetime}-\eqref{eq:condition_spinor} in terms of the original gauge $\Omega_\mu$ and the gauge transformations \eqref{eq:Theta_GR},\eqref{eq:Xi_GR}, we get:
  \begin{align}
	&\Omega_{\tilde\ell} + \tilde\ell^\mu\,\Xi^\nu R_{\nu\mu} + \nabla_{\tilde\ell}\,\Theta = 0 \ ; \label{eq:FHD:gaugeEquation1} \\
	\tilde w^\alpha\partial_\alpha\big(&\Omega_{\bar{m}}+\bar{m}^\mu\,\Xi^\nu R_{\nu\mu} + \nabla_{\bar{m}}\Theta\big) = 0 \ , \label{eq:FHD:gaugeEquation3}
  \end{align}
  where $\nabla_\mu\Theta$ denotes the covariant derivative w.r.t. the full GR-like connection $\omega_\mu$:
  \begin{align}
  	\nabla_\mu\Theta \equiv \partial_\mu\Theta + \{\omega_\mu,\Theta\} \ .
  \end{align}
  Our goal is to solve \eqref{eq:FHD:gaugeEquation1}-\eqref{eq:FHD:gaugeEquation3} for the gradients $\nabla_{\bar m}\Theta\,(= \partial_{\bar m}\Theta)$ and $\partial_{\bar m}\Xi^\mu$ on the shared ray; since $\Theta,\Xi^\mu$ themselves vanish on the shared ray, this will determine the desired component $\tilde\Omega_{\bar{m}}$ in the new gauge. 

  Let us now start by expanding eq. \eqref{eq:FHD:gaugeEquation1} to first order in $h$. Plugging in \eqref{eq:next_x}-\eqref{eq:next_ell}, the $\nabla_{\tilde\ell}\Theta$ piece reads:
\begin{equation}
	\begin{split}
		\nabla_{\tilde\ell}\,\Theta(x;y) ={}& \nabla_{\tilde\ell}\,\Theta\big(x_0 + h\bar m;y\big) \\
		={}& \nabla_{\tilde\ell_0}\Theta(x_0;y) + h\partial_{\bar{m}}\nabla_{\tilde\ell_0}\Theta(x_0;y) \\
		&+ht\nabla_{\bar{m}}\Theta(x_0;y) + h\sigma^\mu_{\alpha\dot\alpha}q^\alpha (\dots)^{\dot{\alpha}}\nabla_\mu\Theta(x_0;y) \ .
	\end{split} \label{eq:nabla_ell_Theta}
\end{equation}
Here, the first term vanishes, since $\Theta$ vanishes on the shared ray. The fourth term, with derivatives along $q^\alpha (\dots)^{\dot{\alpha}}$, also vanishes, because $\Theta$ vanishes along the entire shared left-handed plane. We are left with the second and third terms. In the second term, since $\Theta$ vanishes on the shared ray, we can trade the partial derivative $\partial_{\bar m}$ for a covariant one, and then commute the derivatives:
\begin{equation}
	\partial_{\bar{m}}\nabla_{\tilde\ell_0}\Theta=\nabla_{\bar{m}}\nabla_{\tilde\ell_0}\Theta = \nabla_{\tilde\ell_0}\nabla_{\bar{m}}\Theta + \underbrace{\{R_{\bar m\tilde\ell_0},\Theta\}}_{=0\text{ at }x_0} + \nabla_{[\bar{m},\tilde\ell_0]}\Theta\ .
\end{equation}
Computing the Lie-bracket $[\bar{m},\tilde\ell_0]$, we find a cancellation between the two surviving terms in \eqref{eq:nabla_ell_Theta}, as:
\begin{equation}
	[\bar{m},\tilde\ell_0]=-t\bar{m} \quad\Longrightarrow\quad \nabla_{\tilde\ell}\,\Theta(x;y) = h\nabla_{\tilde\ell_0}\nabla_{\bar{m}}\Theta(x_0;y) \ .
\end{equation}
We now plug this back into the gauge condition \eqref{eq:FHD:gaugeEquation1}. Recalling that $\Omega_\mu$ vanishes when contracted with vectors $\sigma^\mu_{\alpha\dot\alpha}q^\alpha(\dots)^{\dot\alpha}$ along the shared left-handed plane, and that $\Xi^\mu$ vanishes at $x_0$, eq. \eqref{eq:FHD:gaugeEquation1} reduces to (discarding a common factor of $h$):
\begin{equation}
 t\,\Omega_{\bar{m}}(x_0;y) + \tilde\ell_0^\mu\big(\partial_{\bar{m}}\Xi^\nu(x_0;y)\big)R_{\nu\mu}(x_0;y) + \nabla_{\tilde\ell_0}\nabla_{\bar{m}}\Theta(x_0;y) = 0 \ . \label{eq:FHD:gaugeEquation1Expanded}
\end{equation}
We now perform a similar expansion for the second gauge condition \eqref{eq:FHD:gaugeEquation3}. Using the vanishing of $\Theta,\Xi^\mu$ on the shared ray, we get:
\begin{equation}
 \tilde w^\alpha\partial_\alpha\Big(\Omega_{\bar{m}} + h\partial_{\bar{m}}\Omega_{\bar{m}} + h\bar{m}^\mu(\partial_{\bar{m}}\Xi^\nu)R_{\nu\mu} + \nabla_{\bar{m}}\Theta + h\partial_{\bar{m}}\nabla_{\bar{m}}\Theta \Big)=0\ ,
 \label{eq:FHD:gaugeEquation2IntermediateExpansion}
\end{equation}  
where everything is again evaluated at $x_0$. Using our expansion \eqref{eq:next_w} of $\tilde w^\alpha$, together with the property $q^\alpha\partial_\alpha\Omega_\mu = 0$ which holds everywhere in the original gauge, eq. \eqref{eq:FHD:gaugeEquation2IntermediateExpansion} becomes (discarding a common factor of $\sqrt{t}$):
\begin{equation}
	\begin{split}
		hq^\alpha\partial_\alpha&\bigg(\bar{m}^\mu(\partial_{\bar{m}}\Xi^\nu)R_{\nu\mu} + \partial_{\bar{m}}\nabla_{\bar m}\Theta \bigg) \\
		&+\left((1 + O(h))q^\alpha - \frac{h}{2}\,\chi^\alpha \right)\partial_\alpha\nabla_{\bar{m}}\Theta - \frac{h}{2}\,\chi^\alpha\partial_\alpha\Omega_{\bar{m}} = 0 \ .
	\end{split}
\end{equation}
The term with $q^\alpha\partial_\alpha$ acting on $\nabla_{\bar{m}}\Theta$ can be discarded. This is because we're on the shared ray, where $\nabla_\mu\Theta$ defines the difference between the old and new gauges $\tilde\Omega_\mu - \Omega_\mu$, both of which satisfy the gauge condition $q^\alpha\partial_\alpha\tilde\Omega_\mu = q^\alpha\partial_\alpha\Omega_\mu = 0$. We are left with (discarding a common factor of $h$): 
\begin{equation}
  q^\alpha\partial_\alpha\Big(\bar{m}^\mu(\partial_{\bar{m}}\Xi^\nu)R_{\nu\mu} + \partial_{\bar{m}}\nabla_{\bar{m}}\Theta \Big) 
	= \frac{1}{2}\,\chi^\alpha\partial_\alpha\Big(\Omega_{\bar m} + \nabla_{\bar{m}}\Theta\Big) \ . \label{eq:FHD:gaugeEquation2IntermediateExpansion2}
\end{equation}
We can now guess the form of the solution to eqs. \eqref{eq:FHD:gaugeEquation1Expanded},\eqref{eq:FHD:gaugeEquation2IntermediateExpansion2}. We write the following ansatz for the unknown derivatives of $\Theta,\Xi^\mu$ along $\bar m^\mu$, in terms of some functions $f,g_1,g_2$ depending on $t$ and on helicity $u\partial_u - 1$:
\begin{align}
	\nabla_{\bar{m}}\Theta(x_0;y) &= f(t,u\partial_u)\,\Phi_+(x_0;u)\Big\vert_{u = \braket{qy}^2/t} \ ; \label{eq:ansatz_Theta} \\
    \partial_{\bar{m}}\Xi^\mu(x_0;y) &= -\frac{m^\mu}{t}\,\partial_u\left( g_1(t,u\partial_u) -\frac{1}{t}\,g_2(t,u\partial_u)\,\partial_{\tilde\ell_0} \right)\Phi_+(x;u)\Big\vert_{x = x_0; u = \braket{qy}^2/t} \ . \label{eq:ansatz_Xi}
\end{align}
This ansatz immediately gives us some simplifications. Let's focus first on the $\tilde\ell_0^\mu(\partial_{\bar{m}}\Xi^\nu)R_{\nu\mu}$ term in eq. \eqref{eq:FHD:gaugeEquation1Expanded}. In our ansatz \eqref{eq:ansatz_Xi}, we have:
\begin{align}
	\tilde\ell_0^{[\mu}\partial_{\bar{m}}\Xi^{\nu]} \sim \tilde\ell_0^{[\mu} m^{\nu]} \sim q^\alpha q^\beta\sigma^\mu_{\alpha\dot\alpha}\sigma_\beta^{\nu\dot\alpha} \ , \label{eq:ell_Xi}
\end{align} 
which vanishes upon contraction with all the $\Phi_+$-dependent deformation terms in our expression \eqref{eq:R_ansatz} for $R_{\nu\mu}$. This leaves only the pure de Sitter term $R_{\nu\mu}^{\text{dS}}$, and eq. \eqref{eq:FHD:gaugeEquation1Expanded} simplifies to:
\begin{align}
	t\,\Omega_{\bar{m}}(x_0;y) + \tilde\ell_0^\mu\big(\partial_{\bar{m}}\Xi^\nu(x_0;y)\big)R^{\text{dS}}_{\nu\mu}(x_0;y) + \nabla_{\tilde\ell_0}\nabla_{\bar{m}}\Theta(x_0;y) = 0 \ . \label{eq:FHD:gaugeEquation1Linearized}\\ 
\end{align}
Plugging in the lightcone ansatz \eqref{eq:ansatz_omega},\eqref{eq:R_ansatz} and our ansatz \eqref{eq:ansatz_Theta}-\eqref{eq:ansatz_Xi} for the gauge parameters, this becomes:
\begin{align}
	\begin{split}
		&\left(\frac{1}{2} + g_2(t,u\partial_u)u\partial_u - f(t,u\partial_u)\right)\partial_{\tilde\ell_0}\Phi_+(x;u)\Big\vert_{x=x_0;\ u=\braket{qy}^2/t} \\
		&\quad = \left(t g_1(t,u\partial_u)u\partial_u + t^2\partial_tf(t,u\partial_u) - 2t f(t,u\partial_u)u\partial_u \right)\Phi_+(x_0;u)\Big\vert_{u=\braket{qy}^2/t} \ .
	\end{split} \label{eq:FHD:fgEq1}
\end{align}
Turning to our second equation \eqref{eq:FHD:gaugeEquation2IntermediateExpansion2}, we can similarly simplify the $q^\alpha\partial_\alpha\big(\bar{m}^\mu(\partial_{\bar{m}}\Xi^\nu)R_{\nu\mu}\big)$ term. Since the vector $\partial_{\bar{m}}\Xi^\mu\sim m^\mu$ points along the left-handed plane $\sigma^\mu_{\alpha\dot\alpha}q^\alpha(\dots)^{\dot\alpha}$, all contributions to $\bar{m}^\mu(\partial_{\bar{m}}\Xi^\nu)R_{\nu\mu}$ from the $\Phi_+$ deformation in \eqref{eq:R_ansatz} depend on $y^\alpha$ only via $\braket{qy}$. These contributions then vanish when acted on by $q^\alpha\partial_\alpha$, leaving just the de Sitter contribution $q^\alpha\partial_\alpha\big(\bar{m}^\mu(\partial_{\bar{m}}\Xi^\nu)R^{\text{dS}}_{\nu\mu}\big)$. Eq. \eqref{eq:FHD:gaugeEquation2IntermediateExpansion2} then becomes:
\begin{equation}
	q^\alpha\partial_\alpha\!\left(\bar{m}^\mu(\partial_{\bar{m}}\Xi^\nu)R^{\text{dS}}_{\nu\mu}+\partial_{\bar{m}}\nabla_{\bar{m}}\Theta\right) = \frac{1}{2}\,\chi^\alpha\partial_\alpha(\Omega_{\bar{m}}+\nabla_{\bar{m}}\Theta) \ .
	\label{eq:FHD:gaugeEquation2Linearized}
\end{equation}
Frustratingly, this includes a term that depends on $\partial_{\bar{m}}\nabla_{\bar{m}}\Theta(x_0;y)$, i.e. on the value of $\Theta$ at points removed to order $O(h^2)$ from the shared left-handed plane. This is a problem, since solving the gauge-fixing equations to second order in $h$ appears very cumbersome. Furthermore, the second-order equations end up depending on the third order in much the same way. On a pure de Sitter geometry, this tower of equations is tractable, and continues until order $2s-1$ for every spin $s$ in the gauge parameters $\Theta,\Xi^\mu$. For small values of $s$, we managed to solve the full tower, and found that the second-order piece $q^\alpha\partial_\alpha\big(\partial_{\bar{m}}\nabla_{\bar{m}}\Theta\big)$ in \eqref{eq:FHD:gaugeEquation2Linearized} actually \emph{vanishes}. Let us therefore \emph{assume} that it vanishes in the general case; as we'll see, this will reproduce the correct answer from section \ref{sec:comparing_gauges:field_strengths}. Without this term, plugging our ansatz \eqref{eq:ansatz_omega},\eqref{eq:R_ansatz},\eqref{eq:ansatz_Theta}-\eqref{eq:ansatz_Xi} into eq. \eqref{eq:FHD:gaugeEquation2Linearized} gives:
\begin{equation}
	\Big(2tf(t,u\partial_u) - tg_1(t,u\partial_u) + \big(g_2(t,u\partial_u) - 1\big)\partial_{\tilde\ell_0} \Big)\Phi_+(x;u)\big\vert_{x=x_0;\ u=\braket{qy}^2/t}\ .\label{eq:FHD:fgEq2}
\end{equation}
Eqs. \eqref{eq:FHD:fgEq1},\eqref{eq:FHD:fgEq2} can now be solved, giving:
\begin{equation}
	g_1(t,u\partial_u)=2f(t,u\partial_u)\ ;\quad g_2(t,u\partial_u)=1\ ;\quad f(t,u\partial_u)=u\partial_u+\frac{1}{2} \ ,
\end{equation}
Plugging this into \eqref{eq:ansatz_Theta} and recalling the original lightcone gauge \eqref{eq:ansatz_omega}, we obtain the desired field component $\tilde\Omega_{\bar{m}}$ on the shared lightray in the new gauge:
\begin{equation}
	\tilde\Omega_{\bar{m}}(x_0;y) = \Omega_{\bar{m}}(x_0;y)+\nabla_{\bar{m}}\Theta(x_0;y) = \left(-\frac{1}{2t}\,\partial_{\tilde\ell_0}+u\partial_u+\frac{1}{2}\right)\Phi_+(x_0;u)\big\vert_{u=\braket{qy}^2/t} \ . \label{eq:FHD:gaugeTransformSolution}
\end{equation}
We emphasize that this result is unaltered by the interactions: any interaction terms encoded in $\nabla_\mu$ or $R_{\mu\nu}$, as well as any terms coming from spin-2 deformations to the pure de Sitter geometry, have dropped out.

\subsubsection{Comparing to the result of section \ref{sec:comparing_gauges:field_strengths}} \label{sec:comparing_gauges:gaugetransf:comparison}
	
Let's now rewrite the field component \eqref{eq:FHD:gaugeTransformSolution} in the horizon-adapted gauge in a way that will allow comparison to the result of section \ref{sec:comparing_gauges:field_strengths}. This requires finding the analogue $\tilde\Phi_+(x;u)$ in this gauge of the lightcone fields $\Phi_+(x;u)$. From now on, we evaluate everything on the shared lightray, and drop the ``0'' subscript on $x_0^\mu,\tilde\ell_0^\mu$. In the horizon-adapted gauge, the ansatz \eqref{eq:Omega_ansatz_lightcone} expresses the component \eqref{eq:FHD:gaugeTransformSolution} as:
\begin{equation}
	\tilde\Omega_{\bar{m}}(x;y) = -\frac{\tilde t}{2}\,\tilde\ell^\mu\partial_\mu\!\left.\tilde\Phi_+(x;u)\right|_{u=t\braket{qy}^2} \ . \label{eq:Omega_tilde_ansatz_raw}
\end{equation}
Here, we used $\bar m^\mu$ from \eqref{eq:m} and $\tilde w^\alpha = \sqrt{t}\,q^\alpha$ on the shared ray from \eqref{eq:next_w}, and the deformation terms in the vielbein again don't contribute. Now, recall from \eqref{eq:tilde_t} that the
coordinate $\tilde t$ in \eqref{eq:Omega_tilde_ansatz_raw} should be understood as an infinite constant, with its gradient negligible compared to its value. The lightcone field $\tilde\Phi_+(x;u)$ is therefore vanishing, with the finite quantity of interest being the product $\tilde t\,\tilde\Phi_+(x;u)$ (this is the usual situation for lightcone fields in a horizon limit -- see e.g. \cite{Neiman:2023bkq}). We thus rewrite \eqref{eq:Omega_tilde_ansatz_raw} as:
\begin{equation}
	\tilde\Omega_{\bar{m}}(x;y) = -\frac{1}{2}\,\tilde\ell^\mu\partial_\mu\!\left.\left(\tilde t\,\tilde\Phi_+(x;u)\right)\right|_{u=t\braket{qy}^2} \ . \label{eq:Omega_tilde_ansatz}
\end{equation}
Comparing with \eqref{eq:FHD:gaugeTransformSolution} and decomposing into spins $s$ via \eqref{eq:Phi_sigma_decomposition}, we get the relationship between $\Phi_{+s}(x)$ and $\tilde t\,\tilde\Phi_{+s}(x)$ on the shared ray:
\begin{align}
	\tilde\ell^\mu\partial_\mu\!\left(\tilde t\,\tilde\Phi_{+s}(x)\right) = \frac{1}{t^{2s-2}}\left(\frac{1}{t}\,\tilde\ell^\mu\partial_\mu - 2s + 1 \right)\Phi_{+s}(x) \ .
\end{align}
Using $t$ as a coordinate along the shared ray, so that $\tilde\ell^\mu\partial_\mu = t^2(d/dt)$, this can be packaged as:
\begin{align}
	\frac{d}{dt}\!\left(\tilde t\,\tilde\Phi_{+s}\right) = \frac{d}{dt}\frac{\Phi_{+s}}{t^{2s-1}} \ ,
\end{align}
which integrates trivially into a proportionality relation between $\Phi_{+s}$ and $\tilde t\,\tilde\Phi_{+s}$  at the same point:
\begin{align}
	\tilde t\,\tilde\Phi_{+s} = \frac{\Phi_{+s}}{t^{2s-1}} \ .
\end{align}
Recalling \eqref{eq:tilde_t}, we see that this agrees with the result \eqref{eq:Phi_+} of section \ref{sec:comparing_gauges:field_strengths}.

\section{Discussion} \label{sec:discuss}
	
In this paper, we presented a number of results on (GR+YM)-type theories in 4 spacetime dimensions with cosmological constant. In section \ref{sec:GR_YM}, we considered GR+YM theory itself, presenting two novel Lagrangians \eqref{eq:L}-\eqref{eq:V} and \eqref{eq:L_chiral}-\eqref{eq:V_chiral} that employ the same kind of variables in both sectors, playing maximally similar roles. It would be interesting to see whether the double-copy-like structure of these Lagrangians leads to some interesting consequences in scattering problems, such as the usual S-matrix in the $\Lambda\to 0$ limit \cite{Delfino:2012aj}, or de Sitter static-patch amplitudes directly at $\Lambda\neq 0$ \cite{Albrychiewicz2021,Neiman:2023bkq}. In this context, we expect the chiral formulation \eqref{eq:L_chiral}-\eqref{eq:V_chiral} to be superior. This is because it structures the full theory as a perturbative expansion over its self-dual sector \eqref{eq:L_SD}, which corresponds to organizing all scattering amplitudes as perturbations around the MHV sector. 

It would be interesting to extend the formalism of section \ref{sec:GR_YM} to include spin-$\frac{1}{2}$ matter, perhaps by way of supersymmetry. Indeed, arguably the most complete unification of GR and YM occurs in $\mathcal{N}=2$ supergravity. It would be nice to make contact with it. In the context of the Plebanski-type Lagrangian \eqref{eq:L_unpacked_chiral}, some steps in this direction were made in \cite{Capovilla:1991qb,Capovilla:1991kx}, but it appears that new ideas are needed.

In section \ref{sec:covTh}, we considered the higher-spin self-dual version of GR+YM. Our construction shows how interactions of higher-spin fields with total helicity $+1$ (YM-like) and $+2$ (GR-like) can be combined in a single covariant framework, as well as how to descend from this covariant framework into a lightcone gauge. It would be interesting to see if this ``unification'' holds some lessons for the structure of full HS gravity, or at least its chiral sector \cite{Sharapov2022} (whose interactions have \emph{arbitrary positive} total helicity). 

In section \ref{sec:covTh:HS_SD_YM}, we wrote down and solved the lightcone field equations for higher-spin self-dual YM, showing that the structure is a trivial extension of self-dual YM itself. In particular, the equations and their solution are unaffected by the cosmological constant, or, equivalently, by the conformal factor $t$ in Poincare coordinates. In self-dual YM, this is a well-known consequence of conformal symmetry. It would be interesting to understand whether a similar symmetry governs the higher-spin theory as well.

In section \ref{sec:comparing_gauges}, we studied the problem of comparing two lightcone gauges in the higher-spin self-dual theory, which is relevant for the static-patch scattering problem \cite{Albrychiewicz2021,Neiman:2023bkq}. The linearized portion of our analysis should extend directly into full HS gravity. Moreover, a key step in showing that the linearized result persists non-linearly (section \ref{sec:comparing_gauges:field_strengths}), namely the positive total helicity of the interactions, remains valid in the full chiral theory of \cite{Sharapov2022}. On the other hand, another key step in our argument was \emph{causality}, whose status is unclear in the context of HS gravity's higher-derivative interactions. We will take up this problem in our next paper \cite{CubicLightconeScattering}.

Finally, in section \ref{sec:comparing_gauges:gaugetransf}, we studied the explicit gauge transformation between two lightcone gauges. Our construction involved some unjustified ``greedy'' assumptions, i.e. that only the first derivatives of the gauge parameters on the shared lightray are important. Nevertheless, these assumptions led to the correct result, obtained by other means in section \ref{sec:comparing_gauges:field_strengths}. There might be some structure worth exploring behind this happy coincidence.
	
\section*{Acknowledgments}
	
We are grateful to Keith Glennon, Kirill Krasnov and Mirian Tsulaia for discussions. This project was ignited at the 5th Mons Workshop on Higher Spin Gauge Theories. Our work is supported by the Quantum Gravity Unit of Okinawa Institute of Science and Technology Graduate University (OIST).



\begin{thebibliography}{10}
		
		\bibitem{Vasiliev1990}
		M.~A. Vasiliev, {\slshape {Consistent equation for interacting gauge fields of
				all spins in (3+1)-dimensions},}
		\href{http://dx.doi.org/10.1016/0370-2693(90)91400-6}{{\em Phys. Lett. B}
			{\bfseries 243} (1990) 378--382}.
		
		\bibitem{Anninos2011}
		D.~Anninos, T.~Hartman, and A.~Strominger, {\slshape {Higher Spin Realization
				of the dS/CFT Correspondence},}
		\href{http://dx.doi.org/10.1088/1361-6382/34/1/015009}{{\em Class. Quant.
				Grav.} {\bfseries 34} (2017) 015009}, \href{http://arxiv.org/abs/1108.5735}{{
				arXiv:1108.5735~[hep-th]}}.
		
		\bibitem{David2019}
		A.~David, N.~Fischer, and Y.~Neiman, {\slshape {Spinor-helicity variables for
				cosmological horizons in de Sitter space},}
		\href{https://link.aps.org/doi/10.1103/PhysRevD.100.045005}{{\em Phys. Rev.
				D} {\bfseries 100} (Aug, 2019) 045005},
		\href{http://arxiv.org/abs/1906.01058}{{ arXiv:1906.01058~[hep-th]}}.
		
		\bibitem{Albrychiewicz2020}
		E.~Albrychiewicz and Y.~Neiman, {\slshape {Scattering in the static patch of de
				Sitter space},}
		\href{https://link.aps.org/doi/10.1103/PhysRevD.103.065014}{{\em Phys. Rev.
				D} {\bfseries 103} (Mar, 2021) 065014},
		\href{http://arxiv.org/abs/2012.13584}{{ arXiv:2012.13584~[hep-th]}}.
		
		\bibitem{Albrychiewicz2021}
		E.~Albrychiewicz, Y.~Neiman, and M.~Tsulaia, {\slshape {MHV amplitudes and BCFW
				recursion for Yang-Mills theory in the de Sitter static patch},}
		\href{http://dx.doi.org/10.1007/JHEP09(2021)176}{{\em JHEP} {\bfseries 09}
			(2021) 176}, \href{http://arxiv.org/abs/2105.07572}{{
				arXiv:2105.07572~[hep-th]}}.
		
		\bibitem{Neiman_2024}
		Y.~Neiman, {\slshape Self-dual gravity in de {S}itter space: Light-cone ansatz
			and static-patch scattering,}
		\href{https://link.aps.org/doi/10.1103/PhysRevD.109.024039}{{\em Phys. Rev.
				D} {\bfseries 109} (Jan, 2024) 024039},
		\href{http://arxiv.org/abs/2303.17866}{{ arXiv:2303.17866~[gr-qc]}}.
		
		\bibitem{Krasnov2021}
		K.~Krasnov, E.~Skvortsov, and T.~Tran, {\slshape {Actions for self-dual Higher
				Spin Gravities},} \href{http://dx.doi.org/10.1007/JHEP08(2021)076}{{\em JHEP}
			{\bfseries 08} (2021) 076}, \href{http://arxiv.org/abs/2105.12782}{{
				arXiv:2105.12782~[hep-th]}}.
		
		\bibitem{Bern:2008qj}
		Z.~Bern, J.~J.~M. Carrasco, and H.~Johansson, {\slshape {New Relations for
				Gauge-Theory Amplitudes},}
		\href{http://dx.doi.org/10.1103/PhysRevD.78.085011}{{\em Phys. Rev. D}
			{\bfseries 78} (2008) 085011}, \href{http://arxiv.org/abs/0805.3993}{{
				arXiv:0805.3993~[hep-ph]}}.
		
		\bibitem{Bern:2010ue}
		Z.~Bern, J.~J.~M. Carrasco, and H.~Johansson, {\slshape {Perturbative Quantum
				Gravity as a Double Copy of Gauge Theory},}
		\href{http://dx.doi.org/10.1103/PhysRevLett.105.061602}{{\em Phys. Rev.
				Lett.} {\bfseries 105} (2010) 061602},
		\href{http://arxiv.org/abs/1004.0476}{{ arXiv:1004.0476~[hep-th]}}.
		
		\bibitem{Kawai:1985xq}
		H.~Kawai, D.~C. Lewellen, and S.~H.~H. Tye, {\slshape {A Relation Between Tree
				Amplitudes of Closed and Open Strings},}
		\href{http://dx.doi.org/10.1016/0550-3213(86)90362-7}{{\em Nucl. Phys. B}
			{\bfseries 269} (1986) 1--23}.
		
		\bibitem{Ponomarev:2016lrm}
		D.~Ponomarev and E.~D. Skvortsov, {\slshape {Light-Front Higher-Spin Theories
				in Flat Space},} \href{http://dx.doi.org/10.1088/1751-8121/aa56e7}{{\em J.
				Phys. A} {\bfseries 50} (2017) 095401},
		\href{http://arxiv.org/abs/1609.04655}{{ arXiv:1609.04655~[hep-th]}}.
		
		\bibitem{Skvortsov2018}
		E.~Skvortsov, T.~Tran, and M.~Tsulaia, {\slshape {Quantum Chiral Higher Spin
				Gravity},}
		\href{https://link.aps.org/doi/10.1103/PhysRevLett.121.031601}{{\em Phys.
				Rev. Lett.} {\bfseries 121} (Jul, 2018) 031601},
		\href{http://arxiv.org/abs/1805.00048}{{ arXiv:1805.00048~[hep-th]}}.
		
		\bibitem{Sharapov2022a}
		A.~Sharapov, A.~Sharapov, E.~Skvortsov, E.~Skvortsov, A.~Sukhanov, A.~Sukhanov,
		R.~Van~Dongen, and R.~Van~Dongen, {\slshape {Minimal model of Chiral Higher
				Spin Gravity},} \href{http://dx.doi.org/10.1007/JHEP09(2022)134}{{\em JHEP}
			{\bfseries 09} (2022) 134}, \href{http://arxiv.org/abs/2205.07794}{{
				arXiv:2205.07794~[hep-th]}}. [Erratum: JHEP 02, 183 (2023)].
		
		\bibitem{Sharapov2022}
		A.~Sharapov and E.~Skvortsov, {\slshape {Chiral higher spin gravity in (A)dS4
				and secrets of Chern–Simons matter theories},}
		\href{http://dx.doi.org/10.1016/j.nuclphysb.2022.115982}{{\em Nuclear Physics
				B} {\bfseries 985} (Dec., 2022) 115982},
		\href{http://arxiv.org/abs/2205.15293}{{ arXiv:2205.15293~[hep-th]}}.
		
		\bibitem{CubicLightconeScattering}
		J.~Lang and Y.~Neiman, {\slshape {Causality of higher-spin interactions on the
				(A)dS lightcone, with application to the static patch},}.
		
		\bibitem{Metsaev:2018xip}
		R.~R. Metsaev, {\slshape {Light-cone gauge cubic interaction vertices for
				massless fields in AdS(4)},}
		\href{http://dx.doi.org/10.1016/j.nuclphysb.2018.09.021}{{\em Nucl. Phys. B}
			{\bfseries 936} (2018) 320--351}, \href{http://arxiv.org/abs/1807.07542}{{
				arXiv:1807.07542~[hep-th]}}.
		
		\bibitem{Skvortsov2019}
		E.~Skvortsov, {\slshape {Light-Front Bootstrap for Chern-Simons Matter
				Theories},} \href{http://dx.doi.org/10.1007/JHEP06(2019)058}{{\em JHEP}
			{\bfseries 06} (2019) 058}, \href{http://arxiv.org/abs/1811.12333}{{
				arXiv:1811.12333~[hep-th]}}.
		
		\bibitem{Neiman2024}
		Y.~Neiman, {\slshape {Higher-spin self-dual General Relativity: 6d and 4d
				pictures, covariant vs. lightcone},}
		\href{http://dx.doi.org/10.1007/JHEP07(2024)178}{{\em JHEP} {\bfseries 07}
			(2024) 178}, \href{http://arxiv.org/abs/2404.18589}{{
				arXiv:2404.18589~[hep-th]}}.
		
		\bibitem{Armstrong:2020woi}
		C.~Armstrong, A.~E. Lipstein, and J.~Mei, {\slshape {Color/kinematics duality
				in AdS$_{4}$},} \href{http://dx.doi.org/10.1007/JHEP02(2021)194}{{\em JHEP}
			{\bfseries 02} (2021) 194}, \href{http://arxiv.org/abs/2012.02059}{{
				arXiv:2012.02059~[hep-th]}}.
		
		\bibitem{Albayrak:2020fyp}
		S.~Albayrak, S.~Kharel, and D.~Meltzer, {\slshape {On duality of color and
				kinematics in (A)dS momentum space},}
		\href{http://dx.doi.org/10.1007/JHEP03(2021)249}{{\em JHEP} {\bfseries 03}
			(2021) 249}, \href{http://arxiv.org/abs/2012.10460}{{
				arXiv:2012.10460~[hep-th]}}.
		
		\bibitem{Anastasiou:2014qba}
		A.~Anastasiou, L.~Borsten, M.~J. Duff, L.~J. Hughes, and S.~Nagy, {\slshape
		{Yang-Mills origin of gravitational symmetries},}
		\href{http://dx.doi.org/10.1103/PhysRevLett.113.231606}{{\em Phys. Rev. Lett.} {\bfseries 113} (2014) 231606},
		\href{http://arxiv.org/abs/1408.4434}{{ arXiv:1408.4434~[hep-th]}}.

		\bibitem{Anastasiou:2018rdx}
		A.~Anastasiou, L.~Borsten, M.~J. Duff, S.~Nagy, and M.~Zoccali, {\slshape
		{Gravity as Gauge Theory Squared: A Ghost Story},}
		\href{http://dx.doi.org/10.1103/PhysRevLett.121.211601}{{\em Phys. Rev. Lett.} {\bfseries 121} (2018) 211601},
		\href{http://arxiv.org/abs/1807.02486}{{ arXiv:1807.02486~[hep-th]}}.

		\bibitem{Borsten:2021gyl}
		L.~Borsten, H.~Kim, B.~Jur{\v{c}}o, T.~Macrelli, C.~Saemann and M.~Wolf,
		``Tree-level color{\textendash}kinematics duality implies loop-level color{\textendash}kinematics duality up to counterterms,''
		Nucl. Phys. B \textbf{989}, 116144 (2023)
		doi:10.1016/j.nuclphysb.2023.116144
		[arXiv:2108.03030 [hep-th]].

		\bibitem{Borsten:2021hua}
		L.~Borsten, H.~Kim, B.~Jur{\v{c}}o, T.~Macrelli, C.~Saemann, and M.~Wolf,
		{\slshape {Double Copy from Homotopy Algebras},}
		\href{http://dx.doi.org/10.1002/prop.202100075}{{\em Fortsch. Phys.}
		{\bfseries 69} (2021) 2100075}, \href{http://arxiv.org/abs/2102.11390}{{
		arXiv:2102.11390~[hep-th]}}.

		\bibitem{Borsten:2020zgj}
		L.~Borsten, B.~Jur{\v{c}}o, H.~Kim, T.~Macrelli, C.~Saemann, and M.~Wolf,
		{\slshape {Becchi-Rouet-Stora-Tyutin-Lagrangian Double Copy of Yang-Mills Theory},} \href{http://dx.doi.org/10.1103/PhysRevLett.126.191601}{{\em Phys. Rev. Lett.} {\bfseries 126} (2021) 191601},
		\href{http://arxiv.org/abs/2007.13803}{{ arXiv:2007.13803~[hep-th]}}.

\bibitem{Ben-Shahar:2021zww}
M.~Ben-Shahar and H.~Johansson,
``Off-shell color-kinematics duality for Chern-Simons,''
JHEP \textbf{08}, 035 (2022)
doi:10.1007/JHEP08(2022)035
[arXiv:2112.11452 [hep-th]].

\bibitem{Borsten:2023ned}
L.~Borsten, B.~Jurco, H.~Kim, T.~Macrelli, C.~Saemann and M.~Wolf,
``Double Copy From Tensor Products of Metric BV{\ensuremath{\blacksquare}}-Algebras,''
Fortsch. Phys. \textbf{73}, no.1-2, 2300270 (2025)
doi:10.1002/prop.202300270
[arXiv:2307.02563 [hep-th]].

\bibitem{Borsten:2023paw}
L.~Borsten, B.~Jurco, H.~Kim, T.~Macrelli, C.~Saemann, and M.~Wolf, {\slshape
	{Double-copying self-dual Yang-Mills theory to self-dual gravity on twistor
		space},} \href{http://dx.doi.org/10.1007/JHEP11(2023)172}{{\em JHEP}
	{\bfseries 11} (2023) 172}, \href{http://arxiv.org/abs/2307.10383}{{
		arXiv:2307.10383~[hep-th]}}.

\bibitem{Borsten:2022vtg}
L.~Borsten, B.~Jurco, H.~Kim, T.~Macrelli, C.~Saemann and M.~Wolf,
``Kinematic Lie Algebras from Twistor Spaces,''
Phys. Rev. Lett. \textbf{131}, no.4, 041603 (2023)
doi:10.1103/PhysRevLett.131.041603
[arXiv:2211.13261 [hep-th]].

\bibitem{Borsten:2023reb}
L.~Borsten, B.~Jurco, H.~Kim, T.~Macrelli, C.~Saemann, and M.~Wolf, {\slshape
	{Tree-level color-kinematics duality from pure spinor actions},}
\href{http://dx.doi.org/10.1103/PhysRevD.108.126012}{{\em Phys. Rev. D}
	{\bfseries 108} (2023) 126012}, \href{http://arxiv.org/abs/2303.13596}{{
		arXiv:2303.13596~[hep-th]}}.

\bibitem{Ben-Shahar:2021doh}
M.~Ben-Shahar and M.~Guillen,
``10D super-Yang-Mills scattering amplitudes from its pure spinor action,''
JHEP \textbf{12}, 014 (2021)
doi:10.1007/JHEP12(2021)014
[arXiv:2108.11708 [hep-th]].

		\bibitem{Fronsdal1978}
		C.~Fronsdal, {\slshape Massless fields with integer spin,}
		\href{https://link.aps.org/doi/10.1103/PhysRevD.18.3624}{{\em Phys. Rev. D}
			{\bfseries 18} (Nov, 1978) 3624--3629}.
		
		\bibitem{Fronsdal:1978vb}
		C.~Fronsdal, {\slshape {Singletons and Massless, Integral Spin Fields on de
				Sitter Space (Elementary Particles in a Curved Space. 7.},}
		\href{http://dx.doi.org/10.1103/PhysRevD.20.848}{{\em Phys. Rev. D}
			{\bfseries 20} (1979) 848--856}.
		
		\bibitem{Sleight:2016dba}
		C.~Sleight and M.~Taronna, {\slshape {Higher Spin Interactions from Conformal
				Field Theory: The Complete Cubic Couplings},}
		\href{http://dx.doi.org/10.1103/PhysRevLett.116.181602}{{\em Phys. Rev.
				Lett.} {\bfseries 116} (2016) 181602},
		\href{http://arxiv.org/abs/1603.00022}{{ arXiv:1603.00022~[hep-th]}}.
		
		\bibitem{Capovilla:1989ac}
		R.~Capovilla, T.~Jacobson, and J.~Dell, {\slshape {General Relativity Without
				the Metric},} \href{http://dx.doi.org/10.1103/PhysRevLett.63.2325}{{\em Phys.
				Rev. Lett.} {\bfseries 63} (1989) 2325}.
		
		\bibitem{Capovilla:1990qi}
		R.~Capovilla, T.~Jacobson, and J.~Dell, {\slshape {GRAVITATIONAL INSTANTONS AS
				SU(2) GAUGE FIELDS},} \href{http://dx.doi.org/10.1088/0264-9381/7/1/001}{{\em
				Class. Quant. Grav.} {\bfseries 7} (1990) L1--L3}.
		
		\bibitem{Capovilla:1991qb}
		R.~Capovilla, T.~Jacobson, J.~Dell, and L.~J. Mason, {\slshape {Selfdual two
				forms and gravity},} \href{http://dx.doi.org/10.1088/0264-9381/8/1/009}{{\em
				Class. Quant. Grav.} {\bfseries 8} (1991) 41--57}.
		
		\bibitem{Capovilla:1991kx}
		R.~Capovilla, T.~Jacobson, and J.~Dell, {\slshape {A Pure spin connection
				formulation of gravity},}
		\href{http://dx.doi.org/10.1088/0264-9381/8/1/010}{{\em Class. Quant. Grav.}
			{\bfseries 8} (1991) 59--73}.
		
		\bibitem{Krasnov:2011pp}
		K.~Krasnov, {\slshape {Pure Connection Action Principle for General
				Relativity},} \href{http://dx.doi.org/10.1103/PhysRevLett.106.251103}{{\em
				Phys. Rev. Lett.} {\bfseries 106} (2011) 251103},
		\href{http://arxiv.org/abs/1103.4498}{{ arXiv:1103.4498~[gr-qc]}}.
		
		\bibitem{Krasnov:2011up}
		K.~Krasnov, {\slshape {Gravity as a diffeomorphism invariant gauge theory},}
		\href{http://dx.doi.org/10.1103/PhysRevD.84.024034}{{\em Phys. Rev. D}
			{\bfseries 84} (2011) 024034}, \href{http://arxiv.org/abs/1101.4788}{{
				arXiv:1101.4788~[hep-th]}}.
		
		\bibitem{Krasnov2016}
		K.~Krasnov, {\slshape {Self-Dual Gravity},}
		\href{http://dx.doi.org/10.1088/1361-6382/aa65e5}{{\em Class. Quant. Grav.}
			{\bfseries 34} (2017) 095001}, \href{http://arxiv.org/abs/1610.01457}{{
				arXiv:1610.01457~[hep-th]}}.
		
		\bibitem{Krasnov:2017dww}
		K.~Krasnov, {\slshape {Field redefinitions and Plebanski formalism for GR},}
		\href{http://dx.doi.org/10.1088/1361-6382/aac844}{{\em Class. Quant. Grav.}
			{\bfseries 35} (2018) 147001}, \href{http://arxiv.org/abs/1708.07694}{{
				arXiv:1708.07694~[gr-qc]}}.
		
		\bibitem{Jacobson:1987yw}
		T.~Jacobson and L.~Smolin, {\slshape {The Left-Handed Spin Connection as a
				Variable for Canonical Gravity},}
		\href{http://dx.doi.org/10.1016/0370-2693(87)91672-8}{{\em Phys. Lett. B}
			{\bfseries 196} (1987) 39--42}.
		
		\bibitem{Chalmers:1997sg}
		G.~Chalmers and W.~Siegel, {\slshape {Dual formulations of Yang-Mills theory},}
		\href{http://arxiv.org/abs/hep-th/9712191}{{ arXiv:hep-th/9712191}}.
		
		\bibitem{Chalmers:1996rq}
		G.~Chalmers and W.~Siegel, {\slshape {The Selfdual sector of QCD amplitudes},}
		\href{http://dx.doi.org/10.1103/PhysRevD.54.7628}{{\em Phys. Rev. D}
			{\bfseries 54} (1996) 7628--7633},
		\href{http://arxiv.org/abs/hep-th/9606061}{{ arXiv:hep-th/9606061}}.
		
		\bibitem{DePietri:1998hnx}
		R.~De~Pietri and L.~Freidel, {\slshape {so(4) Plebanski action and relativistic
				spin foam model},} \href{http://dx.doi.org/10.1088/0264-9381/16/7/303}{{\em
				Class. Quant. Grav.} {\bfseries 16} (1999) 2187--2196},
		\href{http://arxiv.org/abs/gr-qc/9804071}{{ arXiv:gr-qc/9804071}}.
		
		\bibitem{Freidel:1999rr}
		L.~Freidel, K.~Krasnov, and R.~Puzio, {\slshape {BF description of higher
				dimensional gravity theories},}
		\href{http://dx.doi.org/10.4310/ATMP.1999.v3.n5.a3}{{\em Adv. Theor. Math.
				Phys.} {\bfseries 3} (1999) 1289--1324},
		\href{http://arxiv.org/abs/hep-th/9901069}{{ arXiv:hep-th/9901069}}.
		
		\bibitem{Mitsou:2019nlt}
		E.~Mitsou, {\slshape {Spin connection formulations of real Lorentzian General
				Relativity},} \href{http://dx.doi.org/10.1088/1361-6382/ab00b1}{{\em Class.
				Quant. Grav.} {\bfseries 36} (2019) 045008},
		\href{http://arxiv.org/abs/1901.11312}{{ arXiv:1901.11312~[gr-qc]}}.
		
		\bibitem{Urbantke:1984eb}
		H.~Urbantke, {\slshape {ON INTEGRABILITY PROPERTIES OF SU(2) YANG-MILLS FIELDS.
				I. INFINITESIMAL PART},} \href{http://dx.doi.org/10.1063/1.526402}{{\em J.
				Math. Phys.} {\bfseries 25} (1984) 2321--2324}.
		
		\bibitem{Neiman:2023bkq}
		Y.~Neiman, {\slshape {Self-dual gravity in de Sitter space: Light-cone ansatz
				and static-patch scattering},}
		\href{http://dx.doi.org/10.1103/PhysRevD.109.024039}{{\em Phys. Rev. D}
			{\bfseries 109} (2024) 024039}, \href{http://arxiv.org/abs/2303.17866}{{
				arXiv:2303.17866~[gr-qc]}}.
		
		\bibitem{Bardeen1996}
		W.~A. Bardeen, {\slshape {Self-Dual Yang-Mills Theory, Integrability and
				Multiparton Amplitudes},} \href{http://dx.doi.org/10.1143/ptps.123.1}{{\em
				Progress of Theoretical Physics Supplement} {\bfseries 123} (1996) 1--8}.
		
		\bibitem{Rosly1997}
		A.~A. Rosly and K.~G. Selivanov, {\slshape {On amplitudes in the self-dual
				sector of Yang-Mills theory},}
		\href{http://dx.doi.org/10.1016/s0370-2693(97)00268-2}{{\em Physics Letters
				B} {\bfseries 399} (Apr., 1997) 135--140},
		\href{http://arxiv.org/abs/hep-th/9611101}{{ arXiv:hep-th/9611101}}.
		
		\bibitem{Delfino:2012aj}
		G.~Delfino, K.~Krasnov, and C.~Scarinci, {\slshape {Pure connection formalism
				for gravity: Feynman rules and the graviton-graviton scattering},}
		\href{http://dx.doi.org/10.1007/JHEP03(2015)119}{{\em JHEP} {\bfseries 03}
			(2015) 119}, \href{http://arxiv.org/abs/1210.6215}{{
				arXiv:1210.6215~[hep-th]}}.
		
	\end{thebibliography}

\end{document}